\documentclass[journal,10pt]{IEEEtran}

\usepackage[nocompress]{cite}
\usepackage{url}
\usepackage[pdftex]{graphicx}
\usepackage[cmex10]{amsmath}
\usepackage{algorithm}
\usepackage{algorithmic}
\usepackage{array}
\usepackage[caption=false,font=footnotesize,labelfont=sf,textfont=sf]{subfig}
\usepackage{fixltx2e}
\usepackage{stfloats}
\usepackage{multirow}
\usepackage{color}
\usepackage{amssymb}
\usepackage[english]{babel}
\usepackage{amsthm}
\usepackage{bm}

\usepackage[T1]{fontenc}
\usepackage[numbers,sort&compress]{natbib}
\usepackage{url}

\newtheorem{theorem}{Theorem}
\newtheorem{lemma}{Lemma}

\newtheorem{definition}{Definition}

\newtheorem{observation}{Observation}
\newtheorem{problem}{Problem}

\definecolor{cadmiumgreen}{rgb}{0.0, 0.42, 0.24}

\ifodd 1212

\newcommand{\com}[1]{\textbf{\color{red} (Comment: #1)}}
\newcommand{\comlin}[1]{\textbf{\color{magenta} (Lin Comment: #1)}}
\else

\newcommand{\com}[1]{}
\newcommand{\comlin}[1]{}
\fi

\def\eq{\triangleq}
\def\K{\mathcal{K}}
\def\N{\mathcal{N}}
\def\BB{\mathcal{B}}
\def\LL{\mathcal{L}}
\def\M{\mathcal{M}}

\def\Uhome{U_{\textsc{h}}}
\def\Uroam{U_{\textsc{r}}}
\def\Cserve{C_{\textsc{s}}}

\begin{document}

\title{Incentivizing Wi-Fi Network Crowdsourcing: A Contract Theoretic Approach}

\author{Qian~Ma,~\IEEEmembership{Member,~IEEE,}
		Lin~Gao,~\IEEEmembership{Senior Member,~IEEE,}
        Ya-Feng~Liu,~\IEEEmembership{Member,~IEEE,}
        and~Jianwei~Huang,~\IEEEmembership{Fellow,~IEEE}
        \vspace{-8mm}
\thanks{This work is supported by the General Research Funds (Project Number CUHK 14206315 and CUHK 14219016) established under the University Grant Committee of the Hong Kong Special Administrative Region, China,
and the National Natural Science Foundation of China (Grant Number
61771162, 11631013, and 11688101).}
\thanks{Part of this work has been presented at the 14th International Symposium on Modeling and Optimization in Mobile, Ad Hoc and Wireless Networks (WiOpt), Tempe, Arizona, USA, May 9-13, 2016 \cite{WiOpt}.}
\thanks{Qian Ma and Jianwei Huang are with the Department of
Information Engineering, The Chinese University of Hong Kong. E-mail: \{qianma,\ jwhuang\}@ie.cuhk.edu.hk.
Lin Gao (Corresponding Author) is with the School of Electronic and Information Engineering, Harbin Institute of Technology, Shenzhen, China. E-mail: gaol@hit.edu.cn.
Ya-Feng Liu is with the State Key Laboratory of Scientific and Engineering Computing, Institute of Computational Mathematics and Scientific/Engineering Computing, Academy of Mathematics and Systems Science, Chinese Academy of Sciences, China. E-mail: yafliu@lsec.cc.ac.cn.
}
}

\maketitle

 \begin{abstract}
Crowdsourced wireless community network enables individual users to share their private Wi-Fi access points (APs) with each other, hence can achieve a large Wi-Fi coverage with a small deployment cost via crowdsourcing.
This paper presents a novel \emph{contract-based} incentive framework to incentivize such a Wi-Fi network crowdsourcing under incomplete information (where each user has certain \emph{private} information such as mobility pattern and Wi-Fi access quality).
In the proposed framework, the network operator designs and offers a set of contract items to users, each consisting of a Wi-Fi access price (that a user can charge others for accessing his AP) and a subscription fee (that a user needs to pay the operator for joining the community).
Different from the existing contracts in the literature, in our contract model each user's best choice depends not only on his private information but also on other users' choices.
This greatly complicates the contract design, as the operator needs to analyze the equilibrium choices of all users, rather than the best choice of each single user.
We first derive the feasible contract that guarantees the users' truthful information disclosure based on the equilibrium analysis of user choice,  and then derive the optimal (and feasible) contract that yields the maximal profit for the operator.
Our analysis shows that a user who provides a higher Wi-Fi access quality is more likely to choose a higher Wi-Fi access price and subscription fee, regardless of the user mobility pattern.
Simulation results further show that when increasing the average Wi-Fi access quality of users, the operator can gain more profit, but (counter-intuitively) offer lower Wi-Fi access prices and subscription fees for users.
\end{abstract}


\IEEEpeerreviewmaketitle

\section{Introduction}\label{sec:intro}

\subsection{Background and Motivation}

\IEEEPARstart{W}{i-Fi} technology is playing an increasingly important role in today's wireless communications.
According to Cisco \cite{Cisco}, more than $50\%$ of global mobile data traffic will be carried via Wi-Fi in 2021.
However, each Wi-Fi access point (AP) often has a limited coverage, generally tens of meters \cite{WiFiCoverage}.
Hence, to provide a city-wide or nation-wide Wi-Fi coverage, the network operator needs to deploy a huge number of Wi-Fi APs, which is usually very expensive. {As an example, KT, LG, and SK Telecom in Korea have invested 44 million dollars to provide Wi-Fi access at 10,000 hotspots in Seoul \cite{Investment}.}

On the other hand, in-home Wi-Fi network is becoming more and more popular during the past several years.
According to \cite{WiFiHousehold}, $66\%$ of global households had deployed Wi-Fi APs by 2014, and the percentage is expected to grow to $90\%$ by 2019.
This motivates us to study a new type of Wi-Fi network that relies on aggregating (crowdsourcing) the large number of existing home Wi-Fi APs already deployed by individual users, instead of deploying new Wi-Fi APs.
Such a novel Wi-Fi network based on crowdsourcing is often called \emph{Crowdsourced Wireless Community Network} \cite{communities}.

Specifically, a crowdsourced wireless community network enables a set of individual users, who own private home Wi-Fi APs, to form a community and share their home Wi-Fi APs with each other \cite{communities}.
These AP owners (APOs) are often called \emph{community members}.
By crowdsourcing the massive home Wi-Fi APs, it is possible to achieve a large (e.g., city-wide or nation-wide) Wi-Fi coverage with a small deployment cost.
A successful commercial example of such a   network is FON \cite{FON}, the world largest Wi-Fi community network with more than 20 million members globally.
In practice, however, individual APOs may not know each other personally, hence lack the proper incentive to share Wi-Fi APs with each other \cite{motivation, WiOptFON, TwoPrice, Competition, Qian-1}.
Therefore, one of the   important issues in such a community network is to provide sufficient \emph{economic incentive} for APOs, such that they are willing to join the community and share their Wi-Fi APs with each other.\footnote{Another equally important issue is to guarantee the security and privacy for community members, which can be achieved through proper hardware and software solutions (as did in \cite{FON}), and is beyond the scope of this work.}

In this work, we focus on studying the incentive issue in crowdsourced wireless community networks.
In particular, we consider the network scenario under {incomplete} information, where each APO has certain \emph{private information} (e.g., his mobility pattern and his provided Wi-Fi access quality)
not known by the operator and other APOs.
We aim to design an \emph{incentive mechanism} that
(i) encourages individual APOs to join the community network and share their home Wi-Fi APs properly,
and (ii) offers a considerable revenue for the network operator to manage such a community network.

\vspace{-2mm}

\subsection{Model and Problem Formulation}

\begin{figure}[t]
  \centering
   \includegraphics[width=0.4\textwidth]{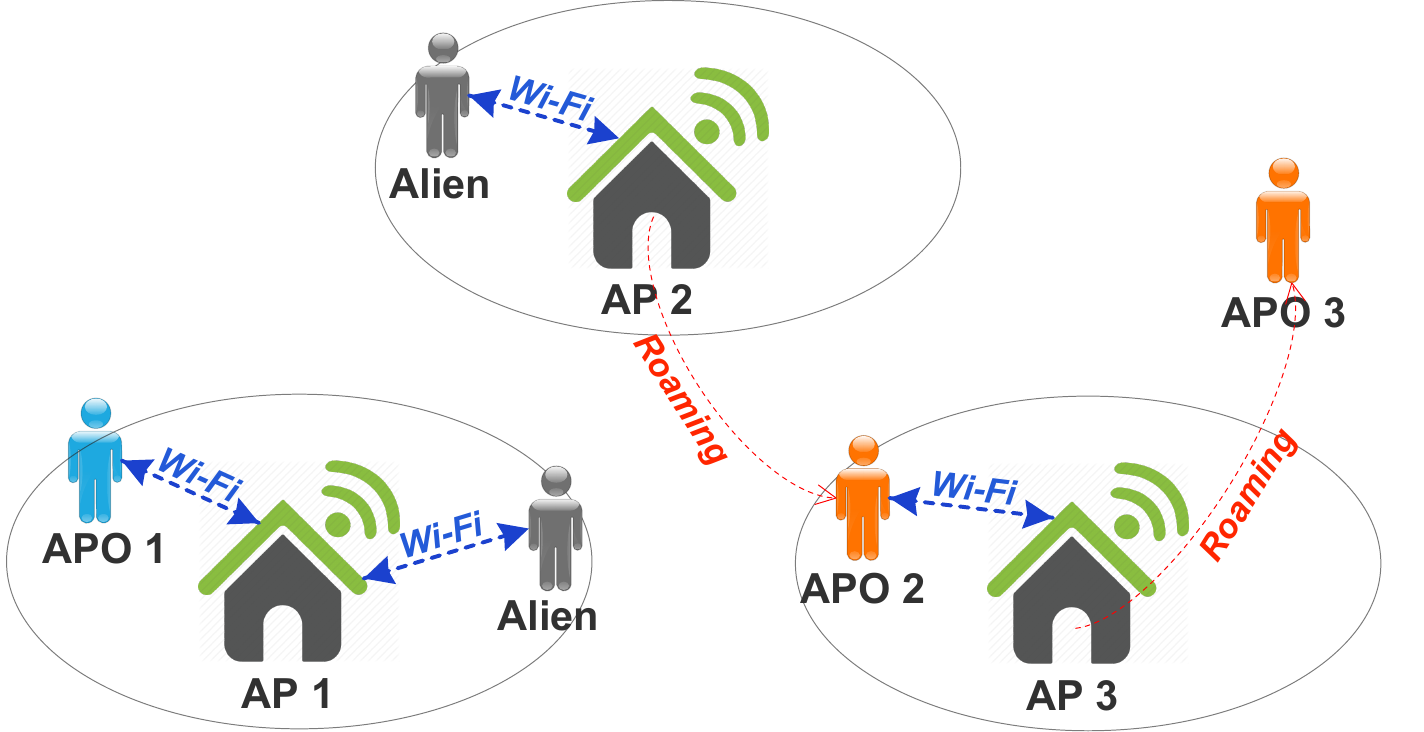}
  \vspace{-2mm}
  \caption{An Example of Crowdsourced Wireless Community Network}\label{fig:model}
  \vspace{-2mm}
\end{figure}

We consider such a crowdsourced wireless community network, where a set of individual APOs form a community and share their Wi-Fi APs with each other.
The community network is open not only to the community members (i.e., APOs) but also to those users (called \emph{Aliens}) who do not own Wi-Fi APs.
Figure \ref{fig:model} illustrates an example of a community network with 3 APOs and 2 Aliens,
where APO 1 stays at home and accesses his own AP,
APO 2 travels to APO 3's home location and accesses AP 3,
APO 3 travels to a location without any member Wi-Fi coverage,
and   Aliens access APs 1 and 2, respectively.

Inspired by the success of FON \cite{FON} and the existing studies \cite{motivation,WiOptFON,Qian-1}, we consider two different sharing schemes for APOs (to share their home Wi-Fi APs), corresponding to two different membership types: \emph{Linus} and \emph{Bill}. Specifically,
\begin{itemize}
  \item As a Linus, an APO can get free Wi-Fi access at other APs in the community, and meanwhile he has to share his own AP without any monetary compensation;
  \item As a Bill, an APO needs to pay for accessing other APs in the community, and meanwhile he can earn some money from sharing his own AP with others.
\end{itemize}
Note that Aliens always need to pay for accessing any AP in the community (as they do not contribute APs).
Thus, from the network operator's perspective, she will get all of the money collected at Linus APs, while only part of the money collected at Bill APs.
Such a dual membership scheme has achieved a great success in practice (e.g., FON \cite{FON}), as it captures two major motivations for APOs to join the community: (i) {getting free Wi-Fi access} and (ii) {earning money}.

In such a community network with dual membership, it is natural to answer the following questions:
\begin{enumerate}
  \item \emph{Membership Selection}: which membership (i.e., Bill or Linus) will each APO select?
  \item \emph{Wi-Fi Pricing}: how to set the Wi-Fi access price (for Bills and Aliens) on each AP?
  \item \emph{Revenue Division}: how to divide the collected revenue at each Bill AP between the   operator and the Bill?\footnote{Note that the operator will get all  revenue collected at a Linus AP.}
\end{enumerate}
Similar as in \cite{motivation,WiOptFON,Qian-1}, we consider such an operation scenario where APOs  make their membership selection decisions and the network operator   makes the Wi-Fi pricing and revenue division decision.
We formulate the interactions between the network operator and APOs as a \emph{two-stage} decision process:
in Stage I, the network operator decides the Wi-Fi pricing and revenue division strategy;
in Stage II, each APO chooses the membership, given the network operator's strategy.

\subsection{Solution and Contributions}

Our focus is to design the  \emph{pricing and revenue division strategy}  for the network operator that provides sufficient   incentive for APOs (to join the community and choose the proper memberships) and meanwhile maximizes her own profit.
The problem is challenging due to several reasons.
First, each APO is associated with certain \emph{private information} (e.g., his mobility pattern and the quality of his provided Wi-Fi access), which cannot be observed by the network operator.
Thus, it is difficult for the network operator to predict the APO's behavior without the complete information.
To this end, some incentive mechanisms are necessary for eliciting the private information of APOs.
Second, APOs' decisions are coupled with each other, as one APO's membership choice will affect other APOs' payoffs, hence affect their membership choices.
Thus, the comprehensive  analysis of APOs' equilibrium choices is challenging, even if the operator's pricing and revenue division strategy are given \cite{Qian-1, WiOptFON, Competition}.

Contract theory \cite{ContractBook} is a promising theoretic tool for dealing with problems with private information, and has been widely adopted in wireless communication  networks \cite{Lin,ContractSpectrum,Lingjie,LingyangSong, Lin-3,Lin-4,
 HanZhu2,HanZhu3}.
Therefore, we propose a contract theoretic framework, in which (i) the network operator offers a \emph{contract} to APOs, which contains a set of \emph{contract items}, each consisting of a Wi-Fi access price and a subscription fee \emph{for Bills};
and (ii) each APO chooses the best membership, and the best contract item if choosing to be Bill.
The key difference between the contract in this work and the existing contracts in \cite{Lin,ContractSpectrum,Lingjie,LingyangSong,Lin-3,Lin-4,
 HanZhu2,HanZhu3} is that in our contract, each APO's best choice depends not only on his private information (as in existing contracts), but also on other APOs' choices.
This greatly complicates the contract design, as the operator needs to analyze the \emph{equilibrium} choices of all APOs, rather than the best choice of each single APO.
In this work, we will first analyze the feasibility of contract and then design the optimal contract systematically.

In summary, the key contributions are summarized below.

\begin{itemize}

\item \emph{Novel Model:}
To our best knowledge, this is the first work that applies the novel contract-based incentive mechanism and   derives the optimal contract to address the incentive issue in crowdsourced wireless community networks.

\item \emph{Novel Theoretical Results:}
The proposed contract mechanism is different from the existing contract mechanisms in the literature due to the coupling of users' choices.
We perform a comprehensive analysis for the users' equilibrium choices, based on which we analyze the feasibility and optimality of   contract systematically.~~~~~~~~~~~~~~~~~~

\item \emph{Practical Insights:}
Our analysis helps us to understand how different users choose their Wi-Fi sharing schemes, which facilitates the network operator to better optimize her profit in different scenarios.
Specifically, we find that (i) a user who provides a higher quality Wi-Fi access is more willing to be a Bill, and choose a larger price and a higher subscription fee regardless of the mobility pattern;
(ii) a user who travels less is more likely to choose to be a Bill to earn money by sharing his Wi-Fi with others;
and (iii) in a network with a larger average Wi-Fi quality, the operator can gain more profit by setting lower prices and subscription fees to all users.

\end{itemize}

The rest of the paper is organized as follows.
In Section \ref{sec:literature}, we review the related works.
In Section \ref{sec:model}, we present the system model.
In Sections \ref{sec:contract}-\ref{sec:OptimalContract}, we provide the contract formulation, analyze the contract feasibility, and design the optimal contract for the homogeneous mobility scenario. 
We provide simulation results in Section \ref{sec:simu} and conclude in Section \ref{sec:conc}.
Due to space limit, we put the detailed proofs and the analysis for the heterogeneous mobility scenario in the online technical report \cite{report}.

\section{Literature Review}\label{sec:literature}

Several recent works have studied the economics and incentive issues in crowdsourced wireless community networks, including the user behavior analysis \cite{Qian-1,motivation,WiOptFON}, the pricing scheme design \cite{TwoPrice}, and the competition analysis \cite{Competition}.
However, all of the above works focused on the scenario of \emph{complete} information, where the network operator is assumed to know the complete information of all APOs.
In practice, however, it is often difficult for the network operator to obtain some APO information, e.g., the APO's  provided Wi-Fi access quality and daily mobility pattern.
In our work, we consider the more practical \emph{incomplete} information scenario.

Contract theory \cite{ContractBook} is a promising theoretic tool for dealing with problems under incomplete information and has been widely-used in supply chain.\footnote{Another important  tool for dealing with problems with incomplete information is auction theory \cite{auction}, which has also been widely-used in wireless networks \cite{auc-1,auc-2,auc-4}.
Auction is more useful for the scenario of allocating limited resources among a set of users with incomplete (asymmetric) information, while contract is more useful for the scenario of motivating different types of users (with private information) behave in a desired way.}
Recently, contract theory has also been introduced to analyze various wireless  network problems, such as spectrum trading/reservation \cite{Lin,Lin-4, Lin-3}, dynamic spectrum sharing \cite{ContractSpectrum,Lingjie}, D2D communications \cite{LingyangSong}, energy harvesting \cite{HanZhu2}, and small-cell caching \cite{HanZhu3}.\footnote{A comprehensive survey of the applications of contract theory in wireless networks can be found in \cite{HanZhu5}.}
{The key difference between our contract and the existing contracts in \cite{Lin,ContractSpectrum,Lingjie,LingyangSong,Lin-3,Lin-4,
 HanZhu2,HanZhu3} is that in our contract, each APO's best choice depends not only on his private information but also on other APOs' choices.
Moreover, in our contract, the private information of each APO is multi-dimensional (i.e., his provided Wi-Fi access quality and his daily mobility pattern), which further complicates the contract design.}

\section{System Model}\label{sec:model}

We consider a set $\N = \{1,...,N\}$ of $N$ individual APOs, each owning a private home Wi-Fi AP, who form a community and share their Wi-Fi APs with each other,
leading to a \emph{crowdsourced wireless community network}.
The community network is managed (coordinated) by a \emph{network operator}, who is responsible for the necessary security, privacy, and incentive issues in such a Wi-Fi network crowdsourcing.
The community network is open not only to APOs but also to a set of $ N_A=a N$ users (called \emph{Aliens}) who do not own Wi-Fi APs, where $a \geq 0$ is the ratio between Aliens and APOs.
Aliens have to pay for accessing the community network (as they do not contribute APs) through, for example, purchasing Wi-Fi passes from the operator.

To ensure the necessary security and privacy, each  AP's channel is divided into two parts: a \emph{private channel} (for the APO's own use) and a \emph{public channel} (for other traveling users' access).
This can be achieved by either specialized hardwares (e.g., customized AP routers as in FON \cite{Channels}) or specific software built in the AP.
Under such a channel division, a traveling user's access to an AP (on the public channel) will not affect the communication of the owner of that particular AP (on the private channel), but will affect other traveling users who access the same AP simultaneously.~~~~~~~~~~~~~~


\emph{1) Wi-Fi Access Quality:}
The Wi-Fi access \emph{quality} provided by different APOs for traveling users (on their public channels) may be different, due to the different wireless characteristics (such as  Wi-Fi standard, backhaul, public channel bandwidth, channel fading, and network congestion) and location popularities.
For example, the Wi-Fi access in a popular location may be more valuable than that in a unpopular location, and hence the APO may choose to install an AP with a carrier grade Wi-Fi standard that provides a better quality than a regular home Wi-Fi.
The Wi-Fi access quality provided by an APO will affect traveling users' lengths of connection on this APO's Wi-Fi, hence affect the APO's potential revenue from sharing his AP (see Section \ref{sec:xxxx} for more details).

\emph{2) APO Type:}
To facilitate the later analysis, we consider a discrete APO (quality) model with a discrete set of $K$ possible Wi-Fi access qualities.
Thus, we can classify all APOs into $K$ \emph{types}, according to the Wi-Fi access qualities that they provide for traveling users (on their public channels).\footnote{Note that
such a discrete quality model can well approximate the continuous quality model by choosing a sufficiently large $K$.}
Namely, the same type APOs will provide the same Wi-Fi access quality.
Let $\theta_k$ denote the Wi-Fi access quality provided by a type-$k$ APO, and   $N_k$ denote the number of type-$k$ APOs.
Without loss of generality, we assume that
$$
\theta_1<\theta_2<\cdots <\theta_K.
$$
It is important to note that the exact type (quality) of each APO is his \emph{private information}, and cannot be observed by the network operator or other APOs.
Similar as in many existing works \cite{Lin,ContractSpectrum,Lingjie,LingyangSong, Lin-3,Lin-4,
 HanZhu2,HanZhu3}, we assume that
the distribution information regarding APO type (i.e., $\{ \theta_k: k \in \K \}$ and $\{N_k: k \in \K\}$, where $\K = \{1,...,K\}$) is public information and known by the network operator and all APOs.
Such distribution information of APO type can be obtained through, for example, historical information or long-term learning.

\emph{3) APO Membership:}
As in \cite{Qian-1, FON,motivation,WiOptFON}, an APO can share his Wi-Fi AP  in two different ways, corresponding to two different membership types: \emph{Linus} and \emph{Bill}. Specifically,
\begin{itemize}

\item
As a Linus, an APO can get free Wi-Fi access at other APs in the community, and meanwhile he has to share his own AP without any monetary compensation.

\item
As a Bill, an APO needs to pay for accessing other APs in the community, and meanwhile he can earn some money from sharing his own AP with others.

\end{itemize}
Moreover, Aliens always need to pay for accessing any AP in the community, as they do not contribute APs.
Thus, from the network operator's perspective, she will get all of the money collected at Linus APs (paid by Bills and Aliens), while only part of the money collected at Bill APs (paid by other Bills and Aliens).
In this work, the network operator will design the Wi-Fi access pricing and revenue sharing (with Bills), and APOs will choose the proper memberships.

Let $p_n$~(\$/min) denote the \emph{Wi-Fi access price} that the network operator designs for AP $n\in \N$, for charging Bills (other than APO $n$) and Aliens who access AP $n$. Then,
\begin{itemize}
\item
When an APO $n$ chooses to be a Bill, he needs to pay for accessing every other AP $m$ at a unit price $p_m$ (per unit Wi-Fi connection time)
and meanwhile he can charge other users (i.e., Bills and Aliens) for accessing his AP at a unit price $p_n$.
\item
When an APO $n$ chooses to be a Linus, he neither pays for accessing other APs, nor charges other users for accessing his AP.
In this case, the network operator will instead charge other users (i.e., Bills and Aliens) for accessing AP $n$.
\end{itemize}

The revenue division between the network operator and each Bill (APO) is via a \emph{subscription fee} that the operator charges the Bill.
Let $\delta_n$ (\$) denote the  {subscription fee} that the network operator charges Bill $n\in \N$.
Note that $\delta_n$ can be negative, which is essentially a bonus from the operator to incentivize  APO $n$ to be a Bill.
Moreover, Linus does not need to pay~any subscription fee.
For clarity, we summarize the properties of different users (i.e., Bill, Linus, and Alien) in Table \ref{table1}.

\begin{table}[t] 
\newcommand{\tabincell}[2]{\begin{tabular}{@{}#1@{}}#2\end{tabular}}
\centering
\caption{A Summary of Three User Types}
\begin{tabular}{|c|c|c|c|}
\hline
\textbf{Type}  & \tabincell{c}{\textbf{Pay for using} \\ \textbf{other APs}} &  \tabincell{c}{\textbf{Paid by sharing} \\ \textbf{his AP}} &
\tabincell{c}{\textbf{Subscription} \\ \textbf{fee}} \\
\hline
Linus  & No & No & No \\
\hline
Bill & Yes & Yes & Yes  \\
\hline
Alien & Yes & N.A. & N.A. \\
\hline
\end{tabular}
\label{table1}
\vspace{-3mm}
\end{table}

\emph{4) APO Mobility:}
To concentrate on the impact of Wi-Fi access quality, in this work we first consider a simple homogeneous mobility pattern for APOs, where (i) \emph{APOs stay at ``home''
with the same probability $\eta$, hence travels outside with probability $1-\eta$}, and (ii) \emph{when traveling outside, each APO travels to every other AP with the same probability $\frac{1-\eta}{N}$.}\footnote{Here we use the approximation  $\frac{1-\eta}{N-1} \approx \frac{1-\eta}{N} $ as $N$ is usually large.}
Moreover, we assume that each Alien travels to every AP with the same probability $\frac{1}{N}$.
Our analysis in Sections \ref{sec:contract}--\ref{sec:OptimalContract} will be based on this homogeneous mobility pattern.

In our online technical report \cite{report}, we will further consider a more general heterogeneous mobility pattern, where different APOs may have different mobility patterns. That is, different APOs may have different probabilities of staying at home (traveling outside).
Note that in this case, an APO's mobility pattern is also his private information.

\section{Contract Formulation}\label{sec:contract}
To deal with the private information, we propose a contract-based operation scheme, where the operator provides a set of \emph{contract items} for each APO to choose.

Specifically, each contract item is the combination of a Wi-Fi access price and a subscription fee, denoted by $\phi \eq (p, \delta)$.
A special contract  item $\phi_0 = (0,0)$ is used to indicate the choice of being Linus.
Based on the \emph{revelation principle} \cite{revalation}, the operator needs to design one contract item for each type of APOs, to induce them to truthfully reveal their types.
Hence a \emph{contract} is such a list of contract items, denoted by
\begin{equation}\label{eq:Phi}
\Phi = \{\phi_k  :  k \in \mathcal{K}\} \eq
\{ (p_k,\delta_k) : k \in \mathcal{K}\},
\end{equation}
where $\phi_k \eq   (p_k,\delta_k) $ denotes the contract item designed for the type-$k$ APOs.
Note that the same type APOs will choose the same contract item, as they have the same parameters, which will be justified later in Section \ref{sec:AP}.

A contract is \emph{feasible} if each APO is willing to choose the contract item designed for his type (i.e., achieving the maximum payoff under the item designed for his type).
A contract is \emph{optimal} if it maximizes the network operator's profit, taking the APOs' truthful information disclosure into consideration.
We will study the feasibility and optimality of contract in Sections \ref{sec:feasibility} and \ref{sec:OptimalContract}, respectively.

 \vspace{-3mm}
\subsection{Operator Profit}

\begin{table}[t] 
\vspace{-2mm}
\newcommand{\tabincell}[2]{\begin{tabular}{@{}#1@{}}#2\end{tabular}}
\centering
\caption{Key Notations}
\begin{tabular}{|c|p{6.5cm}|}
\hline
$\mathcal{K}$  & The set of APO types, $\mathcal{K}=\{1,2,\ldots,K\}$   \\
\hline
$\theta_k$  & The Wi-Fi quality provided by the type-$k$ APOs  \\
\hline
$N_k$  & The number of the type-$k$ APOs  \\
\hline
$p_k$  & The unit Wi-Fi access price at the type-$k$ APs \\
\hline
$\delta_k$  & The subscription fee of the type-$k$ APOs \\
\hline
$\phi_k$ & The contract item designed for the type-$k$ APOs \\
\hline
$\phi_0$ & The contract item chosen by Linus \\
\hline
$\Phi$ & The contract provided by the network operator \\
\hline
$\mathcal{B}(\Phi)$  & The set of APO types choosing Bills under $\Phi$  \\
\hline
$\mathcal{L}(\Phi)$  & The set of APO types choosing Linus under $\Phi$  \\
\hline
$N_B(\Phi)$  & The number of Bills, $N_B(\Phi)=\sum_{k\in\mathcal{B}(\Phi)}N_k$  \\
\hline
$N_L(\Phi)$  & The number of Linus, $N_L(\Phi)=\sum_{k\in\mathcal{L}(\Phi)}N_k$  \\
\hline
$N_A$  & The number of Aliens, $N_A=a  N$  \\
\hline
$a$  & The ratio between Aliens and APOs, $a\geq 0$  \\
\hline
$\eta$  & The probability of an APO staying at home  \\
\hline
 \end{tabular}
\label{table2}
\vspace{-3mm}
\end{table}

We now characterize the network operator's profit under a given \emph{feasible} contract $\Phi= \{\phi_k :  k \in \mathcal{K}\}$, where each APO will choose  the contract item designed for his type.

Let $\LL(\Phi) \eq \{ k \in \K \ | \ \phi_k = \phi_0\}$ denote the set of APO types choosing Linus and $N_L(\Phi)=\sum_{k\in\LL(\Phi)} N_k$ denote the total number of Linus.
Let $\BB(\Phi) \eq \{ k \in \K \ | \ \phi_k \neq \phi_0\}$ denote the set of APO types choosing Bills  and $N_B(\Phi)=\sum_{k\in\BB(\Phi)} N_k$ denote the total number of Bills.
Whenever there is no confusion, we will also write $\LL(\Phi),\BB(\Phi), N_L(\Phi), N_B(\Phi)$ as $\LL,\BB, N_L, N_B$ for notational convenience.
Table \ref{table2} lists the key notations in this paper.

\subsubsection{Profit Achieved from a Bill AP}
If a type-$k$ APO is a Bill who chooses $\phi_k = (p_k, \delta_k) $, the operator's profit achieved from this AP is simply the subscription fee $\delta_k$.

\subsubsection{Profit Achieved from a Linus AP}\label{sec:xxxx}
If a type-$k$ APO is a Linus who chooses $\phi_k = \phi_0$,
the network operator will charge other users (Bills and Aliens) accessing this AP directly, at a price $p_0$ per unit Wi-Fi connection time.
With the assumption of homogeneous mobility, the expected number of Bills accessing this AP is $\sum_{i\in\mathcal{B}}\frac{ 1-\eta }{N}  N_i = \frac{ 1-\eta }{N}  N_B $, and the expected number of Aliens accessing this AP is $\frac{N_A}{N}  = a $.
We further denote the average Wi-Fi connection time of a Bill or Alien on this (type-$k$) AP under the price $p $ as $d_k(p )$.
Intuitively, $d_k(p)$ reflects a Bill's or Alien's demand for using a type-$k$ AP, which is non-negative, monotonically decreasing with the price $p$, and monotonically increasing with quality $\theta_k$.
We will further discuss the demand function at the end of Section \ref{sec:contract}.
Based on the above analysis, the total profit from this type-$k$ Linus AP is:
\vspace{-1mm}
\begin{equation}
\textstyle
\left(\frac{1-\eta}{N} N_B(\Phi)+ a\right)  p_0  d_k(p_0),
\vspace{-1mm}
\end{equation}
where $p_0$ is the price charged by the operator on Linus AP.

\subsubsection{Total Profit}
To summarize, the network operator's profit achieved from all APs can be computed as follows:
\vspace{-2mm}
\begin{equation}\label{eq:totalR}
\textstyle
  \sum_{k\in\mathcal{L}(\Phi)}\left[N_k  \left(\frac{1-\eta}{N} N_B(\Phi)+a\right) p_0d_k(p_0)\right]+\sum_{k\in\mathcal{B}(\Phi)}N_k\delta_k ,
\end{equation}
where the first term is the total profit from all Linus APs and the second term is the total profit from all Bill APs.
 
 \vspace{-3mm}
\subsection{APO Payoff}\label{sec:AP}

We now define the payoff of each APO under a \emph{feasible} contract $\Phi= \{\phi_k :  k \in \mathcal{K}\}$, where each APO
will choose the contract item designed for his type.

\subsubsection{Payoff of a Linus}\label{sec:LinusPayoff}

If a type-$k$ APO chooses to be a Linus (i.e., $\phi_k = \phi_0$), he neither pays for using other APs, nor gains from sharing his own AP.
Hence, his payment is zero, and
his payoff is simply the difference between utility and cost.
Let constant $\Uhome$ denote the utility when staying at home and accessing Wi-Fi through his own AP, constant $\Uroam$ denote the utility when traveling outside and accessing Wi-Fi through other APs, and constant $\Cserve$ denote the cost (e.g., energy, bandwidth) of serving other users.
Note that $\Uhome$, $\Uroam$, and $\Cserve$ do not depend on the APO's membership choice.
Then a type-$k$ Linus APO's payoff is:
\begin{equation}\label{eq:profitLinus}
u_k(\phi_0) = u_k(0,0) = \eta \Uhome + (1-\eta) \Uroam - \Cserve.
\end{equation}

\subsubsection{Payoff of a Bill}

If a type-$k$ APO chooses to be a Bill (i.e., $\phi_k \neq \phi_0$),
his payment consists of (i) the revenue earned at his own AP,
(ii) the payment for accessing other APs,
and
(iii) the subscription fee to the network operator.

First, the revenue earned at his own AP equals the payment of other Bills and Aliens accessing his AP.
The average Wi-Fi connection time of a Bill or Alien on this (type-$k$) AP under the price $p_k$ is $d_k(p_k)$.
The average payment of a Bill or Alien for accessing this AP is
\begin{equation}\label{eq:g}
g_k(p_k) \eq p_k  d_k(p_k),
\end{equation}
which monotonically increases with the Wi-Fi quality $\theta_k$, as the demand $d_k(p_k)$ monotonically increases with $\theta_k$.
The expected number of paying users (including the other $N_k-1$ Bills of type-$k$, all   Bills of other types, and all Aliens)   accessing this AP depends on the given contract $\Phi$, and can be computed as follows:
\begin{equation}\label{eq:alpha}
\omega(\Phi) \eq
\textstyle \frac{1-\eta}{N}(N_B(\Phi)-1)+ \frac{N_A}{N}.
\end{equation}
Note that $\omega(\Phi)$ does not depend on $k$, i.e., same for all Bills.
Thus, the revenue earned at his own AP is $\omega(\Phi) g_k(p_k)$.

Second, the expected payment of this (type-$k$) APO for accessing other APs includes (i) the payment for accessing the other $N_k-1$ Bill APs of type-$k$, (ii) the payment for accessing all Bill APs of other types, and (iii) the payment for accessing Linus APs, which can be calculated as
\begin{align}\label{eq:betak}
\beta_k(\Phi) \eq &\textstyle \frac{1-\eta}{N}\Big[ (N_k-1)g_k(p_k) + \sum_{i \in \mathcal{B}(\Phi),i\neq k}N_ig_i(p_i) \notag \\
&\textstyle+ \sum_{j \in \mathcal{L}(\Phi)}N_jg_j(p_0) \Big] .
\end{align}
Similarly, when there is no confusion, we will also write $\omega$ and $\beta_k$ for $\omega(\Phi)$ and $\beta_k(\Phi)$, respectively.

Third, the subscription fee to the network operator is $\delta_k$.
Hence, a type-$k$ Bill APO's total revenue (when choosing $\phi_k$ under a feasible contract $\Phi$) is
$ \omega(\Phi) g_k(p_k)-\delta_k-\beta_k(\Phi) $.
Based on the definition of $\Uhome$, $\Uroam$, and $\Cserve$ in Section \ref{sec:LinusPayoff}, the type-$k$ Bill APO's payoff is
\begin{equation}\label{eq:profitBill}
\begin{aligned}
u_k(\phi_k;\Phi)=u_k(p_k,\delta_k;\Phi) = & \omega(\Phi) g_k(p_k)-\delta_k-\beta_k(\Phi)
\\
& + \eta  \Uhome + (1-\eta)  \Uroam - \Cserve .
\end{aligned}
\end{equation}
Without loss of generality, in the rest of the paper, we normalize $\eta \Uhome + (1-\eta) \Uroam - \Cserve=0$.

By \eqref{eq:profitBill}, we can see that \emph{the payoff of an APO depends not only on his contract item choice, but also on other APOs' choices.}
For example, the first term in $\omega(\Phi)$ depends on how many APOs choose to be Bills, the second term in $\beta_k(\Phi)$ depends on how other Bills choose prices, and the last term in $\beta_k(\Phi)$ depends on how many APOs choose to be Linus.
Such a strategy coupling makes our contract design problem very different from and more challenging to analyze than traditional contract models in literatures \cite{Lin,ContractSpectrum,Lingjie,LingyangSong, Lin-3,Lin-4,
 HanZhu2,HanZhu3}.

To make the model more practical and meanwhile facilitate the later analysis, we   introduce the following assumptions:
\begin{itemize}
  \item[(a)] $  0 \leq p_k \leq p_{\max}$, $ \forall k \in \mathcal{K}  $;
  \item[(b)] $g_k(p) $ increases with $p \in [0, p_{\max}]$,   $ \forall k \in \mathcal{K}  $;
  \item[(c)] $ g_i(p')-g_i(p)>g_j(p')-g_j(p) $, for any types $i>j$ and prices $p'> p$ (where $p, p' \in [0, p_{\max}]$).
\end{itemize}

Assumption (a) implies that there is a maximum allowable price $p_{\max}$ for Wi-Fi access, which is ofter lower than the corresponding cellular access price.
This is because the Wi-Fi access is usually a complement to the cellular access, and hence a high Wi-Fi access price may drive all APOs to the cellular network when traveling.
Assumption (b) implies that the demand of APOs (on other APs) is inelastic
when $p \in [0, p_{\max}]$.
When Wi-Fi costs so little that users use it like water, increasing the price of Wi-Fi access will not reduce revenue. 
Assumption (c) implies that when increasing the Wi-Fi access price by a small value, the payment of a higher type APO (i.e., with a higher quality) will be increased by a larger value than that of a lower-type APO.
This is ofter referred to as  ``\emph{increasing differences} \cite{IncreasingDifference}'' in economics, and is satisfied for many widely-used demand functions. 
One widely-used demand function that satisfies the above assumptions is:
\begin{equation}\label{eq:demand}
\textstyle
d_k(p)=\frac{T}{1+p/\theta_k}.
\end{equation}
Here $T$ is the length of per subscription period (e.g., one month).\footnote{Function $d_k(p)$ is decreasing convex in $p$. When $p=0$, $d_k(0) = T$, which represents the maximum length of connection time. We may also multiple a constant less than 1 to the demand function, for example, to represent that the users will not connect 24 hours a day even with zero price. Such a constant scaling will not change the analysis results.}
\emph{Note that our analysis is general and does not rely on the specific form of \eqref{eq:demand}.}

\section{Feasibility of Contract}\label{sec:feasibility}

In this section, we will study the sufficient and necessary conditions for a contract to be feasible.
Formally, a contract is feasible, if and only if it satisfies the Incentive Compatibility (IC) and Individual Rationality (IR) constraints.

\begin{definition}[Incentive Compatibility -- IC]
A contract $\Phi= \{\phi_k :  k \in \mathcal{K}\}$ is incentive compatible, if for each type-$k$ APO, he can achieve the maximum payoff when choosing the contract item $\phi_k$ intended for his type $k$, i.e.,
\begin{equation}\label{eq:IC}
u_k(\phi_k; \Phi)\geq u_k(\phi_i; \Phi),~ \forall k, i \in \mathcal{K}.
\end{equation}
\end{definition}

\begin{definition}[Individual Rationality -- IR]
A contract $\Phi= \{\phi_k :  k \in \mathcal{K}\}$ is individual rational,  if for   each type-$k$ APO, he can achieve a non-negative payoff when choosing the item $\phi_k$ intended for his type $k$, i.e.,
\begin{equation}\label{eq:IR}
 u_k(\phi_k; \Phi)\geq 0 ,~ \forall k \in \mathcal{K}.
\end{equation}
\end{definition}

Obviously, if a contract satisfies the IC and IR constraints, each APO will choose the contract item designed for his type.
Hence, we can equivalently say that each APO truthfully reveals his type (private information) to the operator, which forms a Nash equilibrium.

In what follows, we will provide the necessary and sufficient conditions for feasible contracts. Due to space limit, we will put the detailed proofs in the online technical report
\cite{report}.

\vspace{-2mm}

\subsection{Necessary Conditions}

We first show the necessary conditions for a feasible contract.
For  convenience, we rewrite a feasible contract as
 $$
 \Phi= \{(0,0): k \in \LL\}\cup \{(p_k,\delta_k): k \in \BB\},
 $$
where $\LL$ and $\BB$ are the type sets of Linus and Bills, respectively, and both are functions of $\Phi$.\footnote{Here we omit the parameter $\Phi$ for notation simplicity.}
Then, we have: 

\begin{lemma}\label{lemma_p_delta}\label{lemma:1}
If a contract $\Phi=\{(0,0): k \in \LL\}\cup \{(p_k,\delta_k): k \in \BB\}$ is feasible, then
$$
p_i > p_j \Longleftrightarrow \delta_i > \delta_j,\quad \forall i,j \in \BB.
$$
\end{lemma}

\begin{lemma}\label{lemma_p_theta}
If a contract $\Phi=\{(0,0): k \in \LL\}\cup \{(p_k,\delta_k): k \in \BB\}$ is feasible,
  then
  $$\theta_i>\theta_j \Longrightarrow p_i>p_j,\quad \forall i,j\in\BB.$$
\end{lemma}

\begin{lemma}\label{lemma_kc}
If a contract $\Phi=\{(0,0): k \in \LL\}\cup \{(p_k,\delta_k): k \in \BB\}$ is feasible, then there exists a critical APO type $m\in\{1,2,\ldots,K+1\}$, such that $k \in \mathcal{L}$ for all $k < m$ and $k \in \mathcal{B}$ for all $k \geq m$, i.e.,
\begin{equation}
\LL=\{1,2,\ldots,m-1\},\BB=\{m,m+1,\ldots,K\}.
\end{equation}
\end{lemma}

Lemma \ref{lemma:1} shows that in a feasible contract, a larger Wi-Fi access price must correspond to a larger subscription fee. 
Lemma \ref{lemma_p_theta} shows that in a feasible contract, a higher type APO will be designed with a higher Wi-Fi access price.
Lemma \ref{lemma_kc} shows that there exists a critical APO type:
\emph{all APOs with types lower than the critical type will choose to be Linus, and all APOs with types higher than or equal to the critical type will choose to be Bills.}
When all APOs are Bills, we have $m=1$; when all APOs are Linus, we have $m=K+1$.

\subsection{Sufficient and Necessary Conditions}

We now show that the above necessary conditions together are also sufficient for a contract to be feasible. 
For notational convenience, we   denote $\mu(\Phi)$ 
as the expected number of Bills and Aliens accessing a particular AP and $\nu(\Phi)$ 
as the expected payment of a particular APO (for accessing other APs) when choosing to be a Bill. That is, 
$$
\textstyle  \mu(\Phi) \eq \frac{1-\eta}{N}  N_B(\Phi) + \frac{N_A}{N}, ~~~~~~~~~~~~~~~~~~~~~~~~~~~~~
$$
$$
\textstyle
  \nu(\Phi)\eq \frac{1-\eta}{N}  \left(\sum\limits_{i \in \mathcal{B}(\Phi)}N_ig_i(p_i)+\sum\limits_{i \in \mathcal{L}(\Phi)}N_ig_i(p_0) \right) . 
$$
Then, the sufficient and necessary conditions for a feasible contract can be characterized by the following theorem.

\begin{theorem}[Feasible Contract]\label{lemma_Feasibility}
A contract $\Phi=\{\phi_k:  k \in \mathcal{K}\} $ is feasible, if and only if the following conditions hold:
\begin{align}
& \mathcal{L}=\{1,\ldots,m-1\}, \mathcal{B}=\{m ,\ldots,K\},\  m\in \{1,...,K+1\}, \label{eq:con1} \\
& 0 \leq p_0 \leq p_{\max},0 \leq p_i\leq p_k \leq p_{\max},\ \forall k,i \in \mathcal{B},i<k, \label{eq:con2} \\
& \delta_k \geq \underline{\delta}_k \eq \mu(\Phi) g_j(p_k)-\nu(\Phi),\ \forall k \in \mathcal{B}, \forall j \in \mathcal{L}, \label{eq:con3} \\
& \delta_k \leq \bar{\delta}_k \eq \omega(\Phi) g_k(p_k)-\beta_k(\Phi),\ \forall k \in \mathcal{B}, \label{eq:con4} \\
& \omega(\Phi) \left( g_i(p_k)-g_i(p_i) \right) \leq \delta_k-\delta_i \leq \omega(\Phi)\left( g_k(p_k)-g_k(p_i) \right), \notag \\
&~~~~~~~~~~~~~~~~~~~~~~~~~~~~~~~~ \forall k,i \in \mathcal{B},i<k. \label{eq:con5}
\end{align}
\end{theorem}

In the above theorem, the condition \eqref{eq:con3} can be derived from the IC constraints of Linus APOs, i.e., a type-$j$ Linus APO will achieve a non-positive payoff if he chooses any contract item designed for Bills, i.e.,  $(p_k,\delta_k), k \in \BB$.
Similarly, the conditions \eqref{eq:con4} and \eqref{eq:con5} can be derived from the IR and IC constraints of Bill APOs.

\section{Optimal Contract Design}\label{sec:OptimalContract}

In this section, we study the optimal contract that maximizes the network operator's profit.
We focus on finding an optimal contract from the feasible contract set characterized by Theorem \ref{lemma_Feasibility}. 
By the revelation principle \cite{revalation}, the optimal contract within the feasible set is the global optimal contract among all possible (feasible and non-feasible) contracts.

\subsection{Optimal Contract Formulation}

By Theorem \ref{lemma_Feasibility} and  the revelation principle, we can formulate the following contract optimization problem:
\begin{problem}[\textbf{Optimal Contract}]
\begin{align}
& \displaystyle \max ~ \sum_{k\in\mathcal{L}}\left[N_k\left((1-\eta)\frac{N_B}{N}+ a\right)g_k(p_0)\right]+\sum_{k\in\mathcal{B}}N_k\delta_k \notag\\
& ~\mbox{s.t.} ~~~ \eqref{eq:con1},\eqref{eq:con2},\eqref{eq:con3},\eqref{eq:con4},\eqref{eq:con5} \notag \\
& ~\mbox{var:} ~~~ m,\ p_0,\ \{(p_k,\delta_k): k \in \BB\}. \notag
\end{align}
\end{problem}

In Problem 1, $m$ is the critical AP type;
$p_0$ is the Wi-Fi access price on Linus APs, where the network operator will charge Bills or Aliens with this price $p_0$;
and $(p_k,\delta_k),k\in \BB$, are the Wi-Fi access prices and subscription fees for Bills.

Next we define the feasible price assignment, which we will use in later discussions.

\begin{definition}[Feasible Price Assignment]\label{def:fp}
A price assignment $\{p_0,$ $p_k:  k \in \mathcal{B} \}$ is \emph{feasible} if it satisfies constraint \eqref{eq:con2}.
\end{definition}

We will solve Problem 1 in the sequential manner. 
First, we derive the best subscription fees $\{\delta_k:k \in \BB\}$, given the feasible price assignment $\{p_0, p_k:k \in \mathcal{B} \}$ and critical type $m$.
Then, we design an algorithm to compute the optimal price assignment $\{p_0,  p_k:k \in \BB\}$, given the critical type $m$.
Finally, substituting the best subscription fees and price assignment, we search the critical type $m$ for the optimal contract. 

\vspace{-3mm}
\subsection{Optimal Subscription Fee}\label{sec:delta}

We first derive the best subscription fees, given the feasible price assignment and critical type.
The subscription fee optimization problem is given below. 

\begin{problem}[\textbf{Optimal Subscription Fees}]
\begin{align}
& \displaystyle \max ~~ \sum_{k\in\mathcal{B}}N_k\delta_k \notag\\
& ~\mbox{s.t.} ~~~~ \eqref{eq:con3},\eqref{eq:con4},\eqref{eq:con5} \notag \\
& ~\mbox{var:} ~~~ \delta_k,  k \in \mathcal{B}. \notag
\end{align}
\end{problem}

It is easy to see that Problem 2 is a linear programming. The optimal solution is given in the following lemma. 

\begin{lemma}
\label{lemma_optimal_delta}
Given a critical type $m\in\{ 1,\ldots,K+1\}$ and a feasible price assignment $\{p_0,p_k:k \in \mathcal{B}\}$,
the optimal subscription fees $\{{\delta}_k^*:k\in \BB \}$ are given by:
\begin{align}
&{\delta}^*_{m }  =\omega g_{m }(p_{m })-\beta_m , \label{eq:delta_1}
 \\
&{\delta}^*_k  = {\delta}^*_{k-1}+\omega \left( g_k(p_k)-g_k(p_{k-1}) \right) ,\label{eq:delta_k}
\end{align}
for all $ k \in\{ m+1,  ..., K\}$.
\end{lemma}

Note that $\omega$ and $\beta_m$ depends on the critical type $m$ and the price assignment $\{p_0,p_k:k \in \mathcal{B}\}$, and can be calculated by \eqref{eq:alpha} and \eqref{eq:betak}, respectively.
For   convenience, we denote 
$$
s_k= \sum_{i=k}^K N_i, 
\forall k\in\mathcal{B}. 
$$
Then, we can rewrite the network operator's profit from Bill APOs in the following way:
\[
\sum_{k\in\mathcal{B}} N_k\delta_k^* = \sum_{k\in\mathcal{B}}f_k(p_k),
\]
where 
\begin{equation*}
f_k(p_k)=\left\{ 
 \begin{aligned} 
& \textstyle \left( \frac{1-\eta}{N}+\omega \right) s_{k}   g_{k}(p_{k})\notag
 -\omega s_{k+1} g_{k+1}(p_{k}), \ k=m
 \\
& \textstyle  \omega  s_k g_k(p_k)- \omega s_{k+1}g_{k+1}(p_k) ,~~~~~~~ m<k<K
\\
& \textstyle \omega N_k g_k(p_k) , ~~~~~~~~~~~~~~~~~~~~~~~~~~~~~~~~~~ k=K
   \end{aligned}
   \right.
\end{equation*}
It is easy to see that
$f_k(p_k)$ is related to $p_k$ only, while independent of $p_t$ for all $t\neq k$. 


\subsection{Optimal Price Assignment}\label{sec:p}

Next we derive the optimal price assignment, given a fixed critical type $m$.
Substitute the best subscription fees $\{\delta_k^*:k\in \BB \}$ derived in Lemma \ref{lemma_optimal_delta}, we can formulate the following optimal price assignment problem. 

\begin{problem}[\textbf{Optimal Price Assignment}]
\begin{align}
& \displaystyle \max ~~ \sum_{k\in\mathcal{L}}\left[N_k\left((1-\eta)\frac{N_B}{N}+a\right)g_k(p_0)\right]+\sum_{k\in\mathcal{B}}f_k(p_k) \notag\\
& ~\mbox{s.t.} ~~~~  \eqref{eq:con2} \notag \\
& ~\mbox{var:} ~~~ p_0, \ \{p_k:k \in \mathcal{B}\}. \notag
\end{align}
\end{problem}

We denote the solution of Problem 3 as $\{{p}^*_0,{p}^*_k:k\in\mathcal{B}\}$, which depends on the critical type $m$.
Hence, we will also write $\{{p}^*_0,{p}^*_k:k\in\mathcal{B}\}$ as $\{{p}^*_0(m),{p}^*_k(m):k\in\mathcal{B}\}$ when we want to emphasize such dependances.

It is easy to see that the optimal price ${p}^*_0$ for Linus is always the price upper-bound $p_{\max}$, i.e.,
$ {p}^*_0=p_{\max}$.
This is because  $g_k(p_0)$ monotonically increases with $p_0\in [0,p_{\max}]$.
The optimal price assignment $\{ {p}^*_k:k\in\mathcal{B}\}$ for Bills can be solved by the following optimization problem:
\begin{align}
& \max_{p_m,...,p_K} ~~ \sum_{k\in\mathcal{B}}f_k(p_k) \label{eq:max} \\
& ~~\mbox{s.t.}~~~ 0 \leq p_{m }\leq p_{m+1}\leq\cdots \leq p_K \leq p_{\max} . \notag
\end{align}
Since $f_k(p_k)$ only depends on $p_k$, we can compute the maximizer of each $f_k(p_k)$ individually, and check whether the feasible price condition in \eqref{eq:max} is satisfied. 
Let ${p}^\dag_k$ denote the price that maximizes $f_k(p_k)$, i.e.,
\begin{equation}\label{eq:plb}
{p}^\dag_k = \arg \max_{0\leq p_k \leq p_{\max}} f_k(p_k),\ \forall k\in\mathcal{B}.
\end{equation}

Note that if $\{{p}^\dag_k:k\in \mathcal{B}\}$ satisfies the constraint in \eqref{eq:max}, i.e., it is a feasible price assignment, then it is exactly the optimal solution of Problem 3, i.e., $  {p}^*_k = {p}^\dag_k, \forall k\in\mathcal{B}$. 
In some cases, however, $\{{p}^\dag_k:k\in \mathcal{B}\}$ may not satisfy the constraint in \eqref{eq:max}, and hence is not feasible. 
In such cases, we can adopt a dynamic algorithm (similar as that in \cite{Lin}) to adjust the price assignment gradually, such that in each step, at least one infeasible price sub-sequence will be removed.\footnote{An infeasible price sub-sequence refers to a subset of prices that do not satisfy the constraint in \eqref{eq:max}.}
The key idea is to optimize all prices in an infeasible price sub-sequence together, rather than optimizing them individually. 
Due to space limit, we put the detailed algorithm in our online technical report \cite{report}.

\subsection{Optimal Critical Type}

Finally, after obtaining the optimal subscription fees in Section \ref{sec:delta} and the optimal price assignment in Section \ref{sec:p}, we can find the optimal critical type $m^*\in\{ 1,\ldots,K+1\}$ through exhaustive search.
Obviously, we only need to search $K+1$ times for the optimal critical type $m^*$.  


\begin{figure*}[t]
 \centering
\begin{minipage}[t]{0.3 \linewidth}
\centering
\includegraphics[width=1\textwidth]{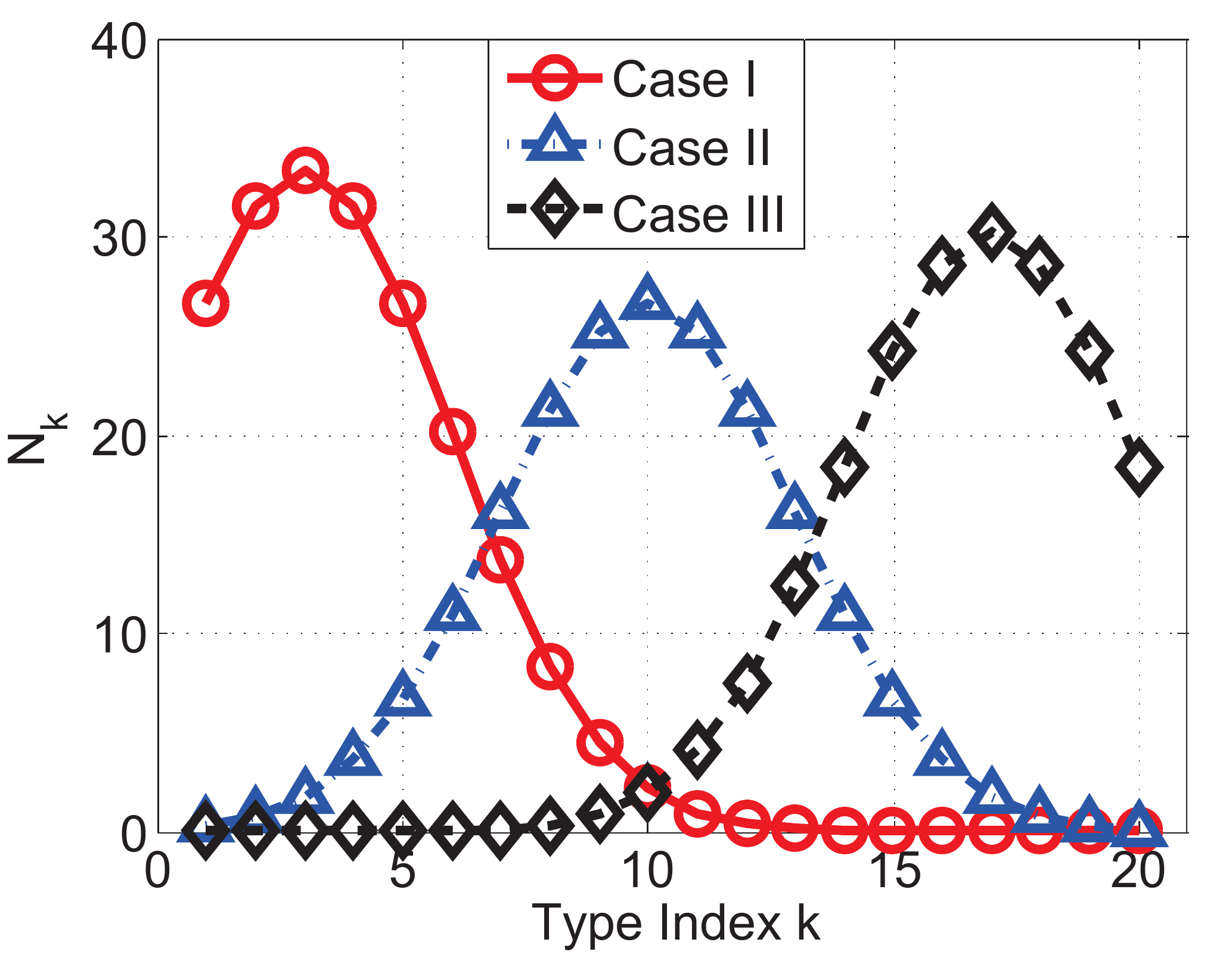}
\vspace{-8mm}  \caption{Distribution of APOs}\label{fig:APD}
\end{minipage}
\begin{minipage}[t]{0.03 \linewidth}
~
\end{minipage}
\begin{minipage}[t]{0.3 \linewidth}
\centering
\includegraphics[width=1\textwidth]{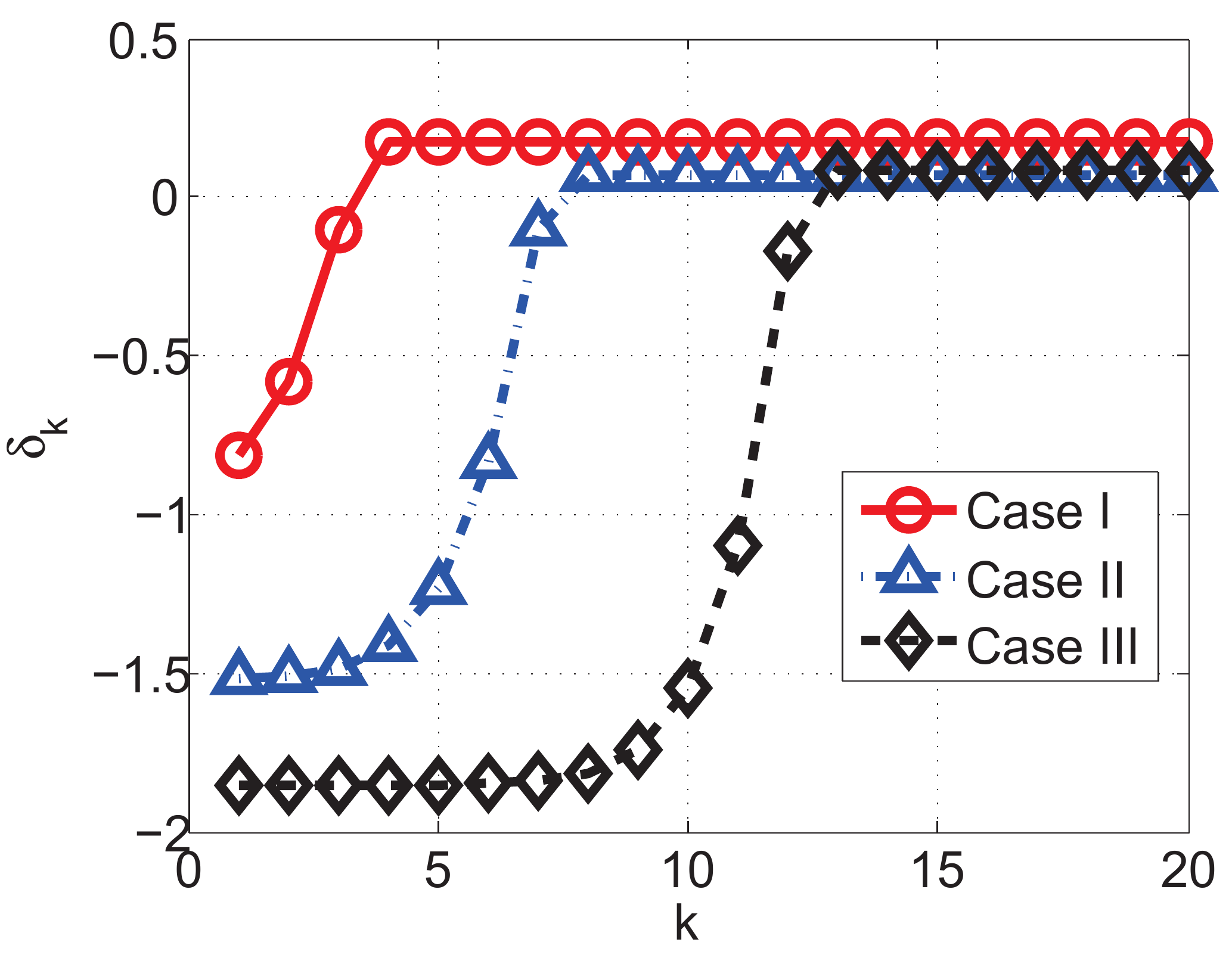}
\vspace{-8mm}  \caption{Subscription Fees for Bills}\label{fig:delta}
\end{minipage}
\begin{minipage}[t]{0.03 \linewidth}
~
\end{minipage}
\begin{minipage}[t]{0.3 \linewidth}
\centering
\includegraphics[width=1\textwidth]{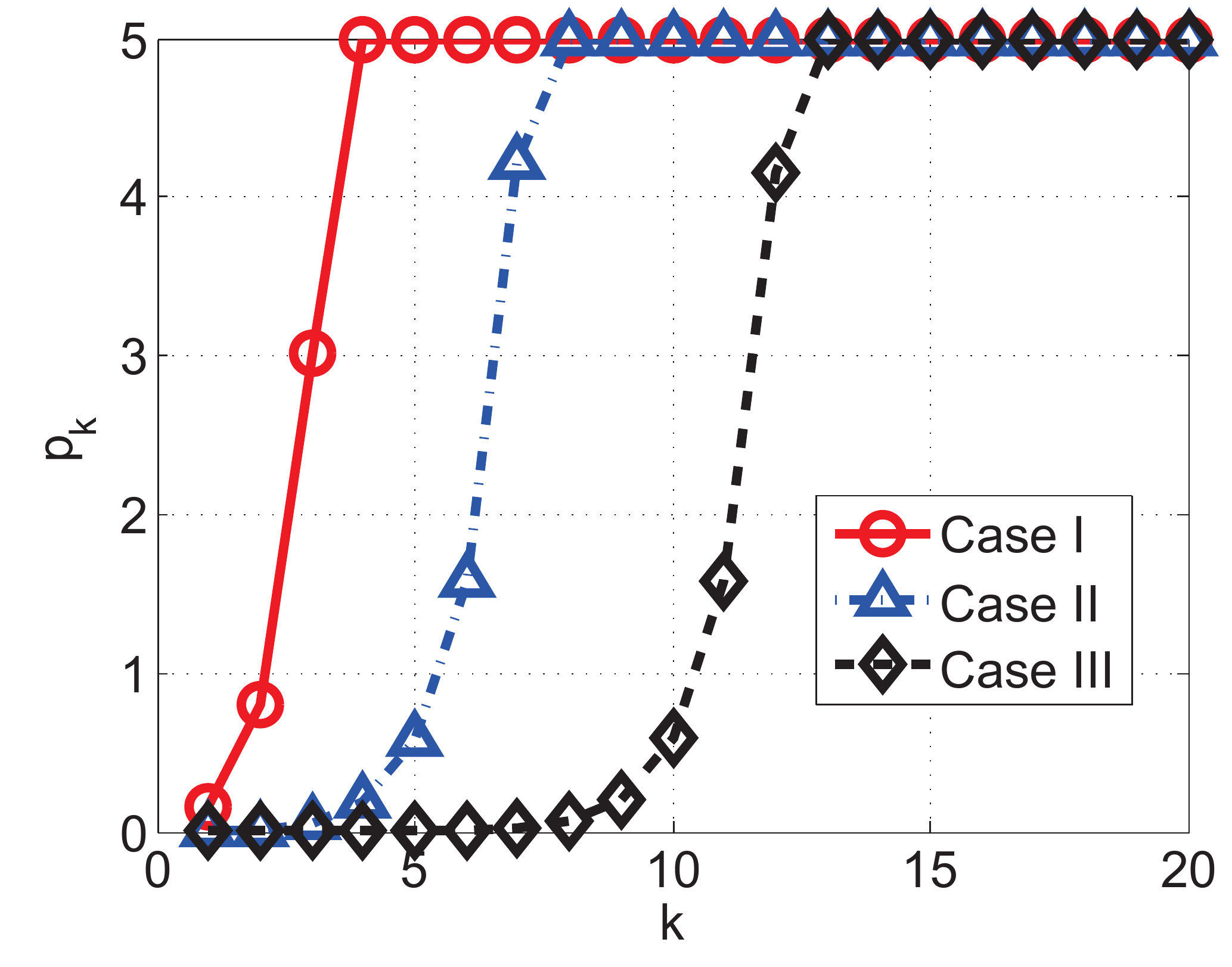}
 \vspace{-8mm} \caption{Wi-Fi Access Prices for Bills}\label{fig:p}
\end{minipage}
\vspace{-3mm}
\end{figure*}

\begin{figure*}[t]
\centering
\begin{minipage}[t]{0.3 \linewidth}
\centering
\includegraphics[width=1\textwidth]{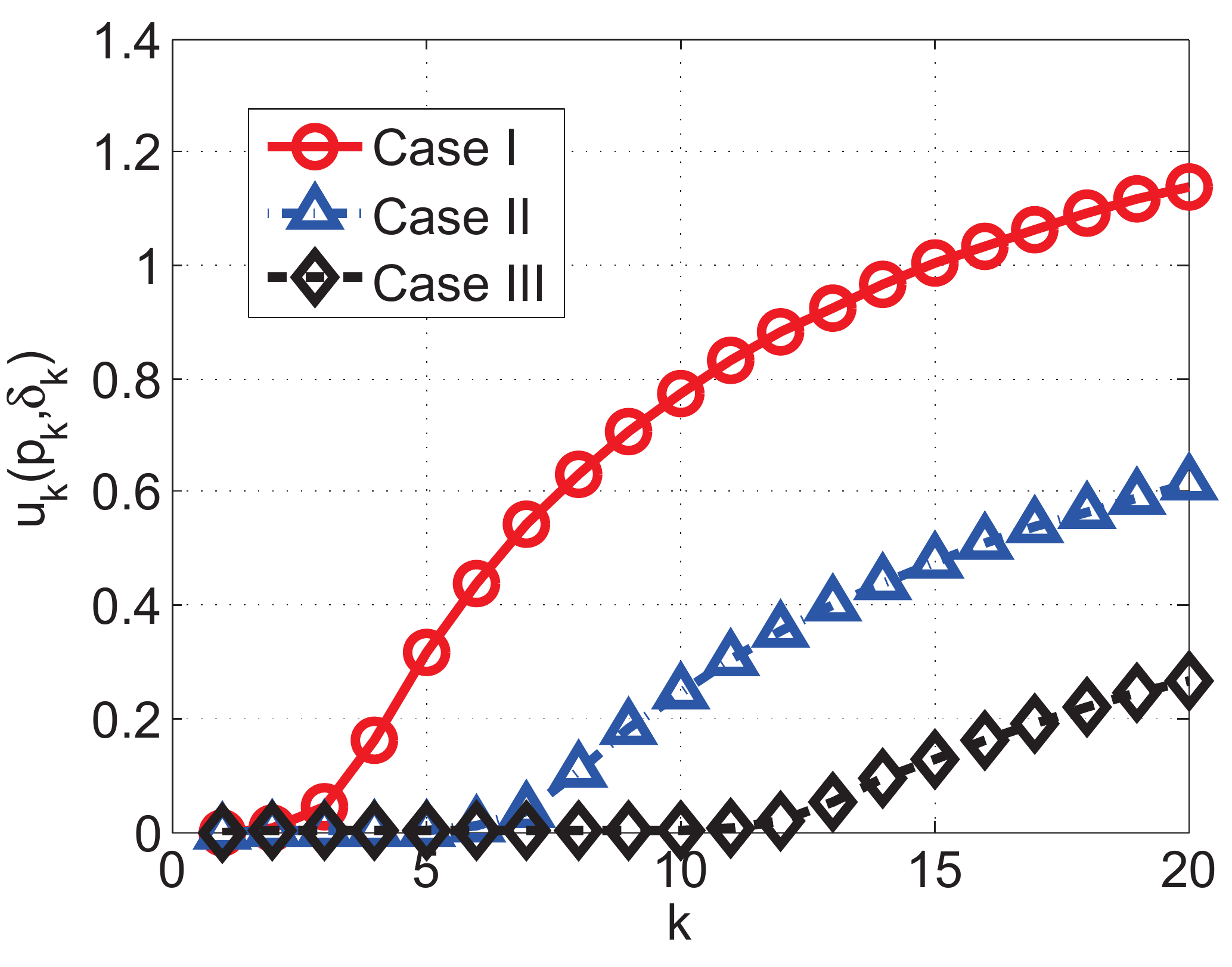}
\vspace{-8mm}  \caption{Payoffs of APOs }\label{fig:OptimalOwnerProfit}
\end{minipage}
\begin{minipage}[t]{0.03 \linewidth}
~
\end{minipage}
\begin{minipage}[t]{0.3 \linewidth}
\centering
\includegraphics[width=1\textwidth]{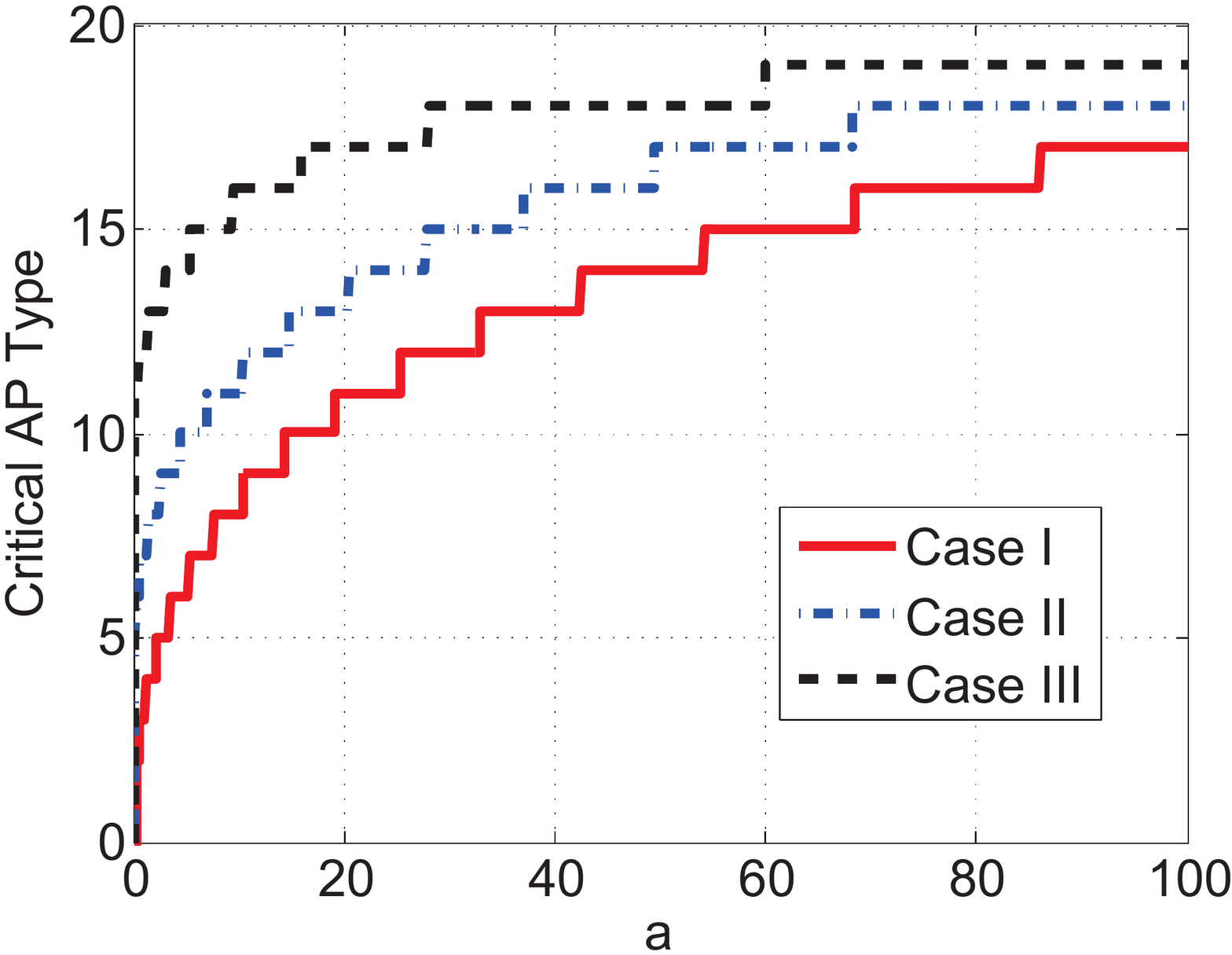}
\vspace{-8mm}  \caption{Critical AP Type}\label{fig:am}
\end{minipage}
\begin{minipage}[t]{0.03 \linewidth}
~
\end{minipage}
\begin{minipage}[t]{0.3 \linewidth}
\centering
\includegraphics[width=1\textwidth]{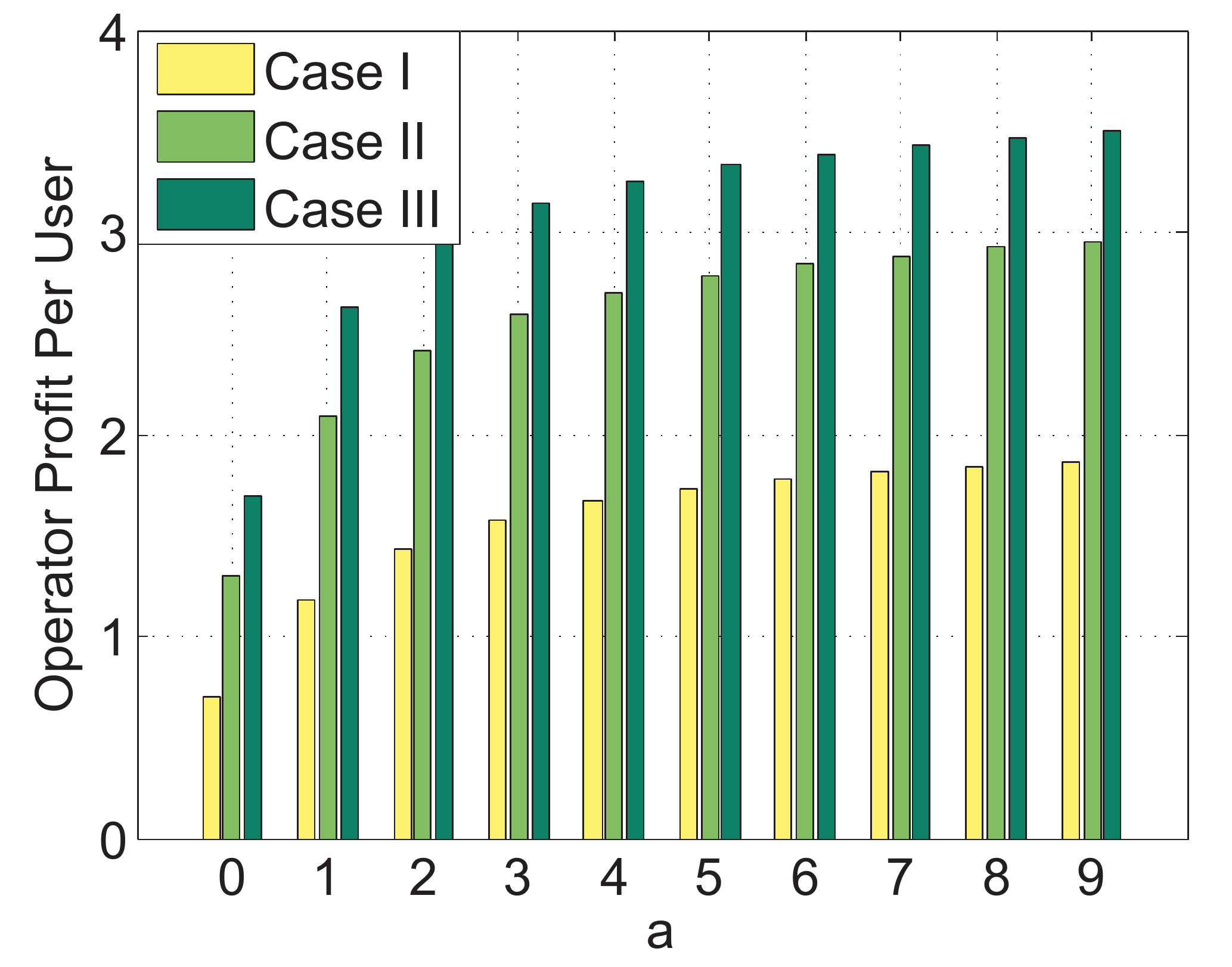}
\vspace{-8mm}  \caption{Network Operator's Profit Per User}\label{fig:OR}
\end{minipage}
\vspace{-3mm}
\end{figure*}

\section{Simulation Results}\label{sec:simu}

In this section, we first simulate a network where all APOs have the homogeneous mobility pattern in Section \ref{sec:simu_homo}, and then simulate a network where APOs have heterogeneous mobility patterns in Section \ref{sec:simu_hete}.\footnote{For more details about the heterogeneous mobility scenario, please refer to our online technical report \cite{report}}

\subsection{Optimal Contract and Impact of Alien Number}\label{sec:simu_homo}

We simulate a network with $N=200$ APOs and $N_A=10$ Aliens. The APOs are classified into $K=20$ types with   Wi-Fi qualities $\boldsymbol{\theta}=\{1,2,\ldots,20\}$.
Each APO stays at home with probability $\eta=0.5$, and travels to other APs with the same probability (i.e., $\frac{ 0.5}{199}$).
The maximum allowable Wi-Fi access price is $p_{\max}=5$.
We study the optimal contracts under three different APO type distributions illustrated in Figure \ref{fig:APD}:
In case I, the population is \emph{low type} APO dominant;
In case II, the population is \emph{medium type} APO dominant;
and In case III, the population is \emph{high type} APO dominant.
Figures \ref{fig:delta} and   \ref{fig:p} illustrate the optimal contract items offered to Bills (i.e., the subscription fees and Wi-Fi access prices) in three cases.
Figure \ref{fig:OptimalOwnerProfit} shows the payoff of each APO in  three cases.

By Figures \ref{fig:delta}--\ref{fig:OptimalOwnerProfit}, we have the following observations for the optimal contracts in the above three cases.

\begin{observation}\label{obs:delta1}
The subscription fee can be negative, in which the operator gives bonus to APOs for sharing Wi-Fi.
Moreover, the subscription fee in Case I (low type APO dominant) is the highest one among all cases (Figure \ref{fig:delta}).
\end{observation}

As a lower type APO charges a smaller price at his own AP but pays higher prices to higher type APOs, his payoff can be negative without proper compensation from the operator.
Hence, the network operator will give bonus to the lower type APO for joining the community and sharing Wi-Fi.
In the low type APO dominant case (Case I), each type APO's payoff as a Bill is higher comparing with other two cases, since lower type APOs charge smaller prices and on average it is less expensive to use other APs.
Hence, the network operator will set higher subscription fees to extract more profit.

\begin{observation}\label{obs:p}
The Wi-Fi access price in Case I (low type APO dominant) is the highest one among all cases (Figure \ref{fig:p}).
\end{observation}

The highest price in Figure \ref{fig:p} corresponds to the highest subscription fee in Figure \ref{fig:delta}.
An APO will choose a higher price since he needs to pay a higher subscription fee.

\begin{observation} \label{obv:APOprofit}
A higher type Bill APO gains more payoff than a lower type Bill APO under all three cases.
Moreover, for each particular APO, it can achieve the largest payoff in Case I (low type APO dominant) among all   cases (Figure \ref{fig:OptimalOwnerProfit}).
\end{observation}

The reason is that the higher type APO receives more revenue on his own AP and pays less to other lower type APOs, hence achieves a  payoff higher than that of a lower type APO.
In the low type APO dominant case (Case I), on average each Bill APO pays less to access other APs than in other two cases.
This turns out to be the dominant factor in determining the payoff, although every Bill APO pays more to the operator (Observation \ref{obs:delta1}) and charges other users more (Observation \ref{obs:p}).


Next, we study how the ratio of Aliens and APOs (i.e., $a=\frac{N_A}{N}$)  affects the optimal contract.
Figure \ref{fig:am} shows the critical AP type $m$ under different values of $a$ in three cases, where we fix $N=200$ and change $N_A$ from $0$ to $20000$ (hence $a$ changes from $0$ to $100$).
Figure \ref{fig:OR} shows the operator's profit per user (including Aliens) under different values of $a$ in three cases, where we fix $N = 200$ and change $N_A$ from $0$ to $1800$ (hence $a={N_A}/{N}$ changes from $0$ to $9$).

By Figures \ref{fig:am} and \ref{fig:OR}, we can obtain the following observations for the optimal contracts in the above three  cases.

\begin{observation} \label{obv:LinusNO}
More APOs will choose to be Linus in Case III (high type APO dominant) than in other cases  (Figure \ref{fig:am}).
\end{observation}

The reason is that higher type APOs charge higher prices (and many of them charge $p_{\max}$).
In the high type APO dominant case (Case III), on average it is more expensive for a Bill to access other APs.
As a result, a lower type APO's payoff as a Bill is smaller due to the higher payment on other APs.
Hence, more APOs will choose to be Linus.

\begin{observation} \label{obv:am}
The number of APOs choosing to be Linus increases with the number of Aliens (Figure \ref{fig:am}).
\end{observation}

Recall the operator's total profit defined in \eqref{eq:totalR}, i.e.,
$$
\textstyle \sum_{k\in\mathcal{L}}\left[N_k\left((1-\eta)\frac{N_B}{N}+a\right)g_k(p_0)\right]+\sum_{k\in\mathcal{B}} N_k \delta_k.
$$
The operator sets the same price on all Linus APs as $p_0=p_{max}$.
As $a$ increases, the term $a g_k(p_{max})$ becomes increasingly important for the operator's total revenue.
Hence, in the optimal contract, the operator increases $m$ to gain more revenue from Aliens' Wi-Fi access on Linus APs.~~~~

\begin{observation} \label{obv:OR}
The network operator's profit per user increases with the number of Aliens. Moreover, given the number of Aliens,  the operator's profit per user in Case III (high type APO dominant) is the largest one among all   cases (Figure \ref{fig:OR}).
\end{observation}

The operator's profit per user increases with $a$, as more Aliens bring more revenue.
The operator's profit is the largest in the high type APO dominant case (Case III).
This is because when the Wi-Fi quality increases, users generate more demand in the network, and hence the operator gains more revenue.

\begin{figure}[t]
\vspace{-5mm}
  \centering
\includegraphics[width=0.35\textwidth]{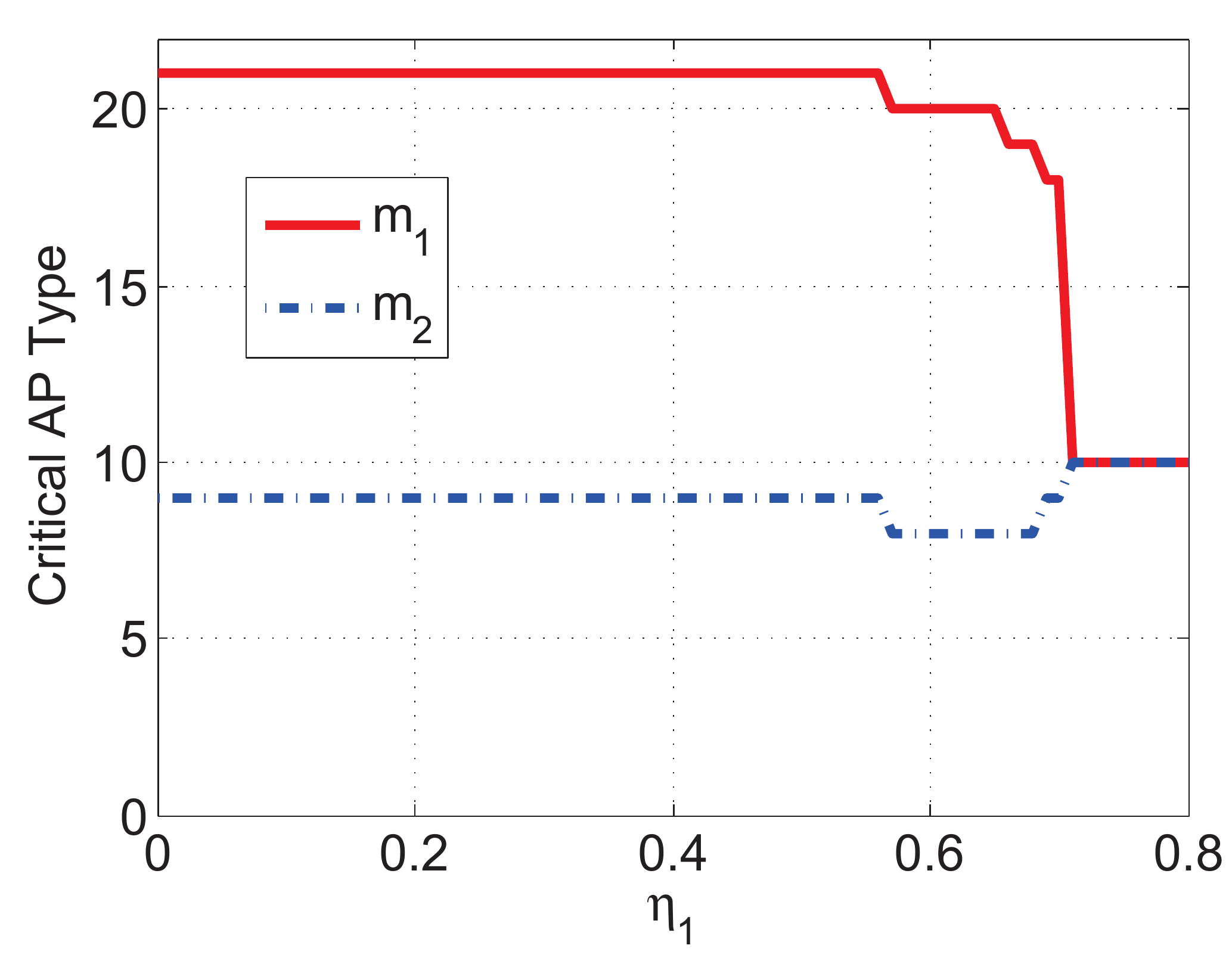}
\vspace{-3mm}
  \caption{Impact of $\eta_1$ to Critical AP Types ($\eta_2=0.8$)}\label{fig:eta1m1m2}
\vspace{-2mm}
\end{figure}

\subsection{Heterogeneous Mobility Scenario}\label{sec:simu_hete}

We now provide simulation results for the heterogeneous mobility scenario. We simulate a network with $N_A=2400$ Aliens and $N=4000$ APOs with $L=2$ mobility patterns ($\eta_1$ and $\eta_2$) and $K=20$ Wi-Fi qualities $\boldsymbol{\theta}=\{1,2,\ldots,20\}$.
The maximum allowable Wi-Fi access price is $p_{\max}=10$.

In our online technical report \cite{report}, we have theoretically proved that if $\eta_1 < \eta_2$, then $m_1 \geq m_2$. Here, we numerically show how the critical AP types $m_1$ and $m_2$ change with the mobility parameters $\boldsymbol{\eta}$.
Figure \ref{fig:eta1m1m2} shows the critical AP types $\{m_1, m_2\}$ under different values of $\eta_1$ when $\eta_2 = 0.8$.

\begin{observation} \label{obv:eta1m1}
The critical AP type $m_1$ decreases with $\eta_1$, i.e., more APOs who stay at home with probability $\eta_1$ choose to be Bills as $\eta_1$ increases (Figure \ref{fig:eta1m1m2}).
\end{observation}

The reason is that as $\eta_1$ increases, APOs with $\eta_1$ stay at home with a larger probability, while travel to other APs with a smaller probability.
So APOs with $\eta_1$ prefer to be Bills to gain profit at his own APs rather than be Linus to enjoy free Wi-Fi at other APs.
Hence, more APOs with $\eta_1$ choose to be Bills as $\eta_1$ increases.

\begin{observation} \label{obv:eta1m2}
(a) The critical AP type $m_2$ first decreases with $\eta_1$, i.e., more APOs who stay at home with probability $\eta_2$ choose to be Bills as $\eta_1$ increases.
(b) When $\eta_1$ is large enough, i.e., $\eta_1 > 0.7$, $m_2$ increases with $\eta_1$, i.e., less APOs who stay at home with probability $\eta_2$ choose to be Bills as $\eta_1$ increases (Figure \ref{fig:eta1m1m2}).
\end{observation}

The reason for Observation \ref{obv:eta1m2}(a) is that as $\eta_1$ increases from small to medium values, more APOs with $\eta_1$ choose to be Bills (Observation \ref{obv:eta1m1}).
Since the number of paying users (Bills) increases, APOs with $\eta_2$ can gain larger profit if they choose to be Bills.
Hence, more APOs with $\eta_2$ choose to be Bills.
The reason for Observation \ref{obv:eta1m2}(b) is that as $\eta_1$ increases from medium to large values, Bill APOs with $\eta_1$ stay at home with a larger probability, so the profit achieved by Bills from these paying users (Bill APOs with $\eta_1$) will decrease.
Hence, less APOs with $\eta_2$ choose to be Bills.

\section{Conclusion}\label{sec:conc}

In this paper, we proposed a novel contract mechanism for crowdsourced wireless community networks under incomplete information.
Different from existing contract mechanisms in the literature, the proposed contract considers the coupling among users' contract item choices, hence is more complicated to design.
We analyzed the feasibility and optimality of the proposed contract systematically based on the user equilibrium analysis.
We also provided simulation results to illustrate the optimal contract and the profit gain of the operator.
Our analysis helps us to understand how different users choose their Wi-Fi sharing schemes, which facilitates the network operator to better optimize her profit in different network and information scenarios.
As for the future work, it is important to study a more general model with heterogeneous users, where different users may have different traffic demands and mobility patterns.
It is also interesting to study the competition among multiple network operators.


\begin{IEEEbiography}[{\includegraphics[width=1in,clip,keepaspectratio]{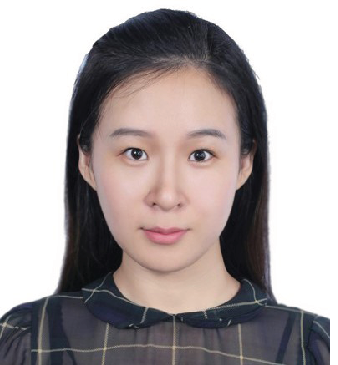}}]{Qian Ma}
 (M'18) is a postdoc researcher in the Department of Information Engineering at The Chinese University of Hong Kong. She received the B.Sc. degree in 2012 from Beijing University of Posts and Telecommunications, Beijing, China, and the Ph.D. degree in 2017 from The Chinese University of Hong Kong, Hong Kong, China. During her Ph.D. study, she visited Professor Tamer Basar at University of Illinois at Urbana-Champaign from September 2015 to February 2016. Her research interests lie in the field of wireless communications and network economics. She is the recipient of the Best Student Paper Award from the IEEE International Symposium on Modeling and Optimization in Mobile, Ad Hoc and Wireless Networks (WiOpt) in 2015. 
\end{IEEEbiography}

\begin{IEEEbiography}[{\includegraphics[width=1in,clip,keepaspectratio]{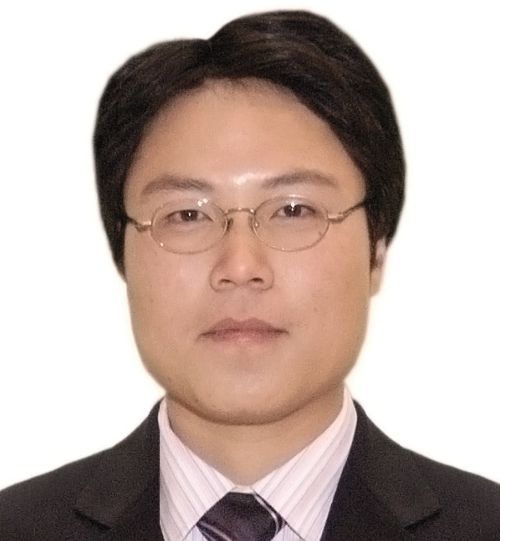}}]{Lin Gao}
(S'08-M'10-SM'16) is an Associate Professor
with the School of Electronic and Information
Engineering, Harbin Institute of Technology,
Shenzhen, China.
He received Ph.D. degree 
in Electronic Engineering from Shanghai Jiao
Tong University in 2010 and served as a Postdoc Fellow in The Chinese University of Hong Kong from 2010 to 2015.
His main research interests 
are in the area of network economics and
games, with applications in wireless communications
and networking. 
He is the co-receipt of 3 Best Paper Awards from WiOpt 2013, 2014, 2015, and 1 Best Paper Award Finalist from IEEE INFOCOM 2016.
He is a receipt of the IEEE ComSoc Asia-Pacific Outstanding Young Researcher Award in 2016.
\end{IEEEbiography}

\begin{IEEEbiography}[{\includegraphics[width=1in,clip,keepaspectratio]{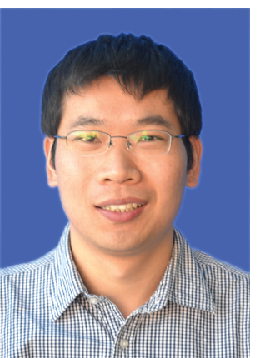}}]{Ya-Feng Liu}(M'12) received the B.Sc. degree in applied mathematics in 2007 from Xidian University, Xi'an, China, and the Ph.D. degree in computational mathematics in 2012 from the Chinese Academy of Sciences (CAS), Beijing, China. During his Ph.D. study, he was supported by the Academy of Mathematics and Systems Science (AMSS), CAS, to visit Professor Zhi-Quan (Tom) Luo at the University of Minnesota (Twins Cities) from February 2011 to February 2012. After his graduation, he joined the Institute of Computational Mathematics and Scientific/Engineering Computing, AMSS, CAS, Beijing, China, in July 2012, where he is currently an Assistant Professor. His main research interests are nonlinear optimization and its applications to signal processing, wireless communications, and machine learning. He is especially interested in designing efficient algorithms for optimization problems arising from the above applications.
Dr. Liu has served as a guest editor of the Journal of Global Optimization. He is a recipient of the Best Paper Award from the IEEE International Conference on Communications (ICC) in 2011 and the Best Student Paper Award from the International Symposium on Modeling and Optimization in Mobile, Ad Hoc and Wireless Networks (WiOpt) in 2015.
\end{IEEEbiography}

\begin{IEEEbiography}[{\includegraphics[width=1in,clip,keepaspectratio]{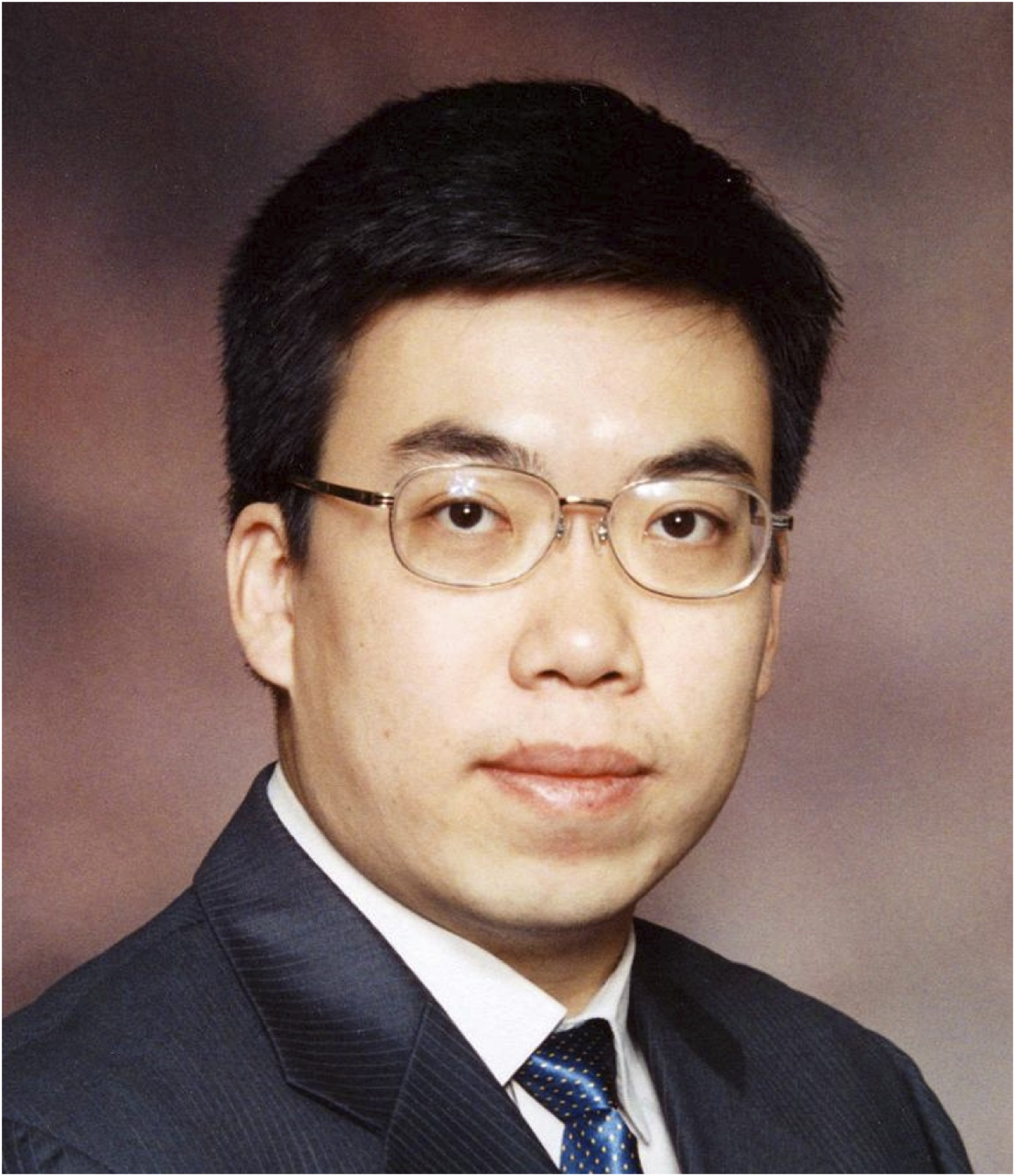}}]{Jianwei Huang}
(F'16) is a Professor in the Department of Information Engineering at The Chinese University of Hong Kong. He is the co-author of 9 Best Paper Awards, including IEEE Marconi Prize Paper Award in Wireless Communications 2011. He has co-authored six books, including the textbook on ``Wireless Network Pricing''. He has served as the Chair of IEEE TCCN and MMTC. He is an IEEE ComSoc Distinguished Lecturer and a Clarivate Analytics Highly Cited Researcher.
\end{IEEEbiography}

\newpage

\appendices

\section{Algorithms for Optimal Price Assignment}

As discussed in Section \ref{sec:p}, the optimal price assignment $\{ {p}^*_k:k\in\mathcal{B}\}$ for \eqref{eq:max} can be obtained by solving each price individually according to \eqref{eq:plb}, as long as the solutions of \eqref{eq:plb} $\{ {p}^\dag_k:k\in\mathcal{B}\}$ satisfy the constraint in \eqref{eq:max}.
In the general case, however, $\{ {p}^\dag_k:k\in\mathcal{B}\}$ may not satisfy the constraint in \eqref{eq:max}, and  hence may not be feasible. 

In this section, we propose a two-stage algorithm to solve the optimal price assignment $\{ {p}^*_k:k\in\mathcal{B}\}$ for \eqref{eq:max}. 
In Stage I, we use a \emph{Dual Algorithm} to solve \eqref{eq:max}, where $\{ {p}^\dag_k:k\in\mathcal{B}\}$ computed in \eqref{eq:plb} will serve  as the initial price choices.
If the solution returned by the Dual Algorithm is feasible, then it is optimal to \eqref{eq:max};
otherwise, we use it as an initial point for a \emph{Dynamic Algorithm} in Stage II, which will return a feasible (but sub-optimal) solution of \eqref{eq:max}.

\subsection{Dual Algorithm}

\begin{algorithm}[b]
\caption{Dual Algorithm}
\label{algo:DualAlgorithm}
\begin{algorithmic}[1]
\REQUIRE
$\{{p}^\dag_k:k \in \mathcal{B}\}$
\ENSURE
$ \{p_k^L:k \in \mathcal{B}\}$

\STATE Initiate $\boldsymbol{\lambda}^{0}=\boldsymbol{0},\boldsymbol{\lambda}^{-1}=\boldsymbol{-1},t=0,\{p_k^t:k \in \mathcal{B}\}= \{{p}^\dag_k:k \in \mathcal{B}\}$\;
\WHILE {If $\exists ~i$ such that $|\lambda_i^t-\lambda_i^{t-1}| > \epsilon$}
\STATE For all $k=m ,\ldots ,K-1$, update
\begin{equation}\label{uplam}
\lambda_k^{t+1}=\max \{ \lambda_k^t+\frac{1}{\sqrt{t+1}} (p_k^t-p_{k+1}^t),0 \}.
\end{equation}
\STATE Calculate $\boldsymbol{p}^{t+1} = \arg \min_{0 \leq \boldsymbol{p} \leq p_{\max}} L(\boldsymbol{p},\boldsymbol{\lambda}^{t+1})$.
\STATE Set $t=t+1$.
\ENDWHILE
\STATE Set $\boldsymbol{\lambda}^L=\boldsymbol{\lambda}^t,\boldsymbol{p}^L = \boldsymbol{p}^t$.
\end{algorithmic}
\end{algorithm}

Let $\lambda_k$ be the Lagrange multiplier associated with the linear constraint $p_k \leq p_{k+1},  k=m ,m+1,\ldots,K-1$.
We denote $\boldsymbol{p}\eq \{p_k:k\in\mathcal{B}\}$, and $\boldsymbol{\lambda}\eq \{\lambda_k:k=m,\cdots,K-1\}$.
Then the Lagrangian of problem \eqref{eq:max} is:
\begin{align}\label{eq:Lagrangian}
L(\boldsymbol{p},\boldsymbol{\lambda}) &= -\sum_{k\in\mathcal{B}}f_k(p_k)+\sum_{k=m }^{K-1}\lambda_k(p_k-p_{k+1}), 
\end{align}
and it can be written as
$$
L(\boldsymbol{p},\boldsymbol{\lambda}) \eq \sum_{k\in\mathcal{B}}L_k(p_k,\boldsymbol{\lambda}),
$$
where $L_k(p_k,\boldsymbol{\lambda})$ only involves $p_k$, and is defined as:
\begin{align}
& L_m(p_m,\boldsymbol{\lambda})=-f_{m }(p_{m })+\lambda_{m }p_{m }, \\
& L_k(p_k,\boldsymbol{\lambda})=-f_k(p_k)+(\lambda_k-\lambda_{k-1})p_k, m < k < K, \\
& L_K(p_K,\boldsymbol{\lambda})=-f_K(p_K)-\lambda_{K-1}p_K.
\end{align}

The dual algorithm aims to solve the convex but possibly nondifferentiable dual problem of \eqref{eq:max}:
\begin{equation}\label{eq:Dual}
\max_{\boldsymbol{\lambda}\geq \boldsymbol{0}} ~d(\boldsymbol{\lambda}) ,
\end{equation}
where
\begin{equation}
d(\boldsymbol{\lambda}) \eq \sum_{k\in\mathcal{B}} \min_{0\leq p_k \leq p_{\max} }L_k(p_k,\boldsymbol{\lambda}). \label{dual2}
\end{equation}
Computing $d(\boldsymbol{\lambda})$ involves $K-m+1$ separate subproblems, $\min_{0\leq p_k \leq p_{\max} }L_k(p_k,\boldsymbol{\lambda}), k \in \BB$, where each subproblem is a single variable minimization problem and is easy to solve.

Note that the right-hand side of \eqref{dual2} might have multiple solutions, hence the dual function $d(\boldsymbol{\lambda})$ may not be differentiable.
We apply the \emph{subgradient method}  to solve the dual problem \eqref{eq:Dual}, i.e., we update the dual variable according to \eqref{uplam} in Algorithm \ref{algo:DualAlgorithm}, where $t$ is the iteration index.
The update rule \eqref{uplam} is intuitive: \emph{the dual variable $\lambda_k$ increases if the constraint $p_k \leq p_{k+1}$ is violated}.
In Algorithm \ref{algo:DualAlgorithm}, $\epsilon \in (0,1)$ is a tolerance parameter.

The sequence  $\{\boldsymbol{\lambda}^t\}$ generated by Algorithm \ref{algo:DualAlgorithm} converges to an optimal dual solution.
Let $\boldsymbol{\lambda}^L$ be the limiting point of the sequences of $\{\boldsymbol{\lambda}^t\}$, and
\begin{equation}
\boldsymbol{p}^L = \arg \sum_{k\in\mathcal{B}} \min_{0\leq p_k \leq p_{\max} }L_k(p_k,\boldsymbol{\lambda}^L) . \label{optp}
\end{equation}
By the weak duality theory, we have
\begin{equation}\label{eq:UB}
 \sum_{k\in\mathcal{B}} f_k(p_k^L) \geq \sum_{k\in\mathcal{B}} f_k( {p}^*_k).
\end{equation}
Recall that $\{ {p}^*_k:k\in\mathcal{B}\}$ is the optimal solution of Problem 3.
Hence, if $\{p_k^L:k \in \mathcal{B}\}$ is feasible to problem \eqref{eq:max}, it is the optimal solution to problem \eqref{eq:max}, i.e.,
$
\{ {p}^*_k:k \in \mathcal{B}\}=\{p_k^L:k \in \mathcal{B}\}.
$
If $\{p_k^L:k \in \mathcal{B}\}$ is not feasible to problem \eqref{eq:max}, we will further compute a feasible solution using the dynamic algorithm introduced later.

To summarize, the dual algorithm fully takes advantage of the separable structure of the objective function in problem \eqref{eq:max};
it is easy to implement, since the optimal solution of the Lagrangian can be obtained easily;
it provides   an upper bound of problem \eqref{eq:max};
it is possible to return   the global solution of problem \eqref{eq:max} with a certificate (if the returned solution is feasible);
if not, it returns   a good approximate solution, which can be used as an initial point for local searching.

\subsection{Dynamic Algorithm}

\begin{algorithm}[b]
\caption{Dynamic Algorithm}
\label{algo:DA}
\begin{algorithmic}[1]
\REQUIRE
$\{p_k^L:k \in \mathcal{B}\}$
\ENSURE
$ \{p_k^D:k \in \mathcal{B}\}$

\STATE Initiate $\{p_k^D:k \in \mathcal{B}\}= \{p_k^L:k \in \mathcal{B}\}$\;
\WHILE {$\{p_k^D:k \in \mathcal{B}\}$ is infeasible}
\STATE Find the first and shortest infeasible sub-sequence $\{p_i^D,p_{i+1}^D,\ldots ,p_j^D\}$ of $\{{p}_k^D, k \in \mathcal{B}\}$.
\STATE Set $\displaystyle p_k^D=\arg \max_{0 \leq p \leq p_{\max}}\sum_{t=i}^jf_t(p),~\forall k=i,\ldots ,j$.
\ENDWHILE
\end{algorithmic}
\end{algorithm}

Now we deal with the case where $\{p_k^L:k \in \mathcal{B}\}$ is not feasible.

\begin{definition}[Infeasible Sub-sequence]\label{def:ifss}
We denote a sub-sequence of $\{p_k:k \in \mathcal{B}\}$, say $\{p_i,p_{i+1},\ldots,p_j\}$, as an infeasible sub-sequence, if $p_i \geq p_{i+1} \geq \cdots \geq p_j$ and $p_i>p_j$.
\end{definition}

We propose to use a Dynamic Algorithm in Algorithm \ref{algo:DA} (based on a similar algorithm proposed in \cite{Lin}) to make all the infeasible sub-sequences in $\{p_k^L:k \in \mathcal{B}\}$ feasible, by enforcing all prices in the infeasible sub-sequence to be equal (Line 4 of Algorithm \ref{algo:DA}).
Specifically, for an infeasible sub-sequence $\{p_i,p_{i+1},\ldots,p_j\}$, we combine $f_i(p_i),f_{i+1}(p_{i+1}),\ldots,f_j(p_j)$ and select an identical price for $p_i,p_{i+1},\cdots,p_j$ by:
\begin{equation}\label{eq:combineEqs}
p_i=p_{i+1}=\cdots=p_j= \arg \max_{0 \leq p \leq p_{\max}}\sum_{t=i}^jf_t(p).
\end{equation}
It is easy to solve \eqref{eq:combineEqs}, as it has only a single variable $p$.
We denote the solution returned by Algorithm \ref{algo:DA} as $ \{p_k^D:k \in \mathcal{B}\}$.

The Dynamic Algorithm will terminate within $K-m$ times, as we combine at least two different $f_i(\cdot)$ at each time,
hence we need to do at most $K-m$ times.

Since $\{p_k^D:k \in \mathcal{B}\}$ satisfies the constraints of problem \eqref{eq:max}, it provides a lower bound of problem \eqref{eq:max}, i.e.,
 \begin{equation}\label{eq:LB}
 \sum_{k\in\mathcal{B}} f_k(p_k^D) \leq \sum_{k\in\mathcal{B}} f_k( {p}^*_k).
 \end{equation}

\subsection{Complexity Analysis}

We now analyze the complexity of the above algorithms.
Specifically, we first need to derive the best subscription fees $\{\delta_k:k \in \BB\}$, given the feasible price assignment $\{p_0,$ $p_k:k \in \mathcal{B} \}$ and the critical type $m$.
Note that the closed-form expressions of the best subscription fees are given in (18) and (19), and hence can be computed without involving complicated optimization.

Then, we adopt a two-stage algorithm to solve the price assignment $\{p_0,\  p_k:k \in \BB\}$, given the critical type $m$.
The complexity of the Dual Algorithm in Stage I is $K \cdot \mathcal{O}\left( {1}/{\epsilon^2} \right)$.
Specifically, the complexity to calculate the dual objective function value and the corresponding subgradient value once is $K$.
The stopping criterion of the Dual Algorithm is that the difference between the objectives of two consecutive iterations is no larger than the tolerance parameter $\epsilon$.
The complexity of such iterations is $\mathcal{O}\left( {1}/{\epsilon^2} \right)$ [30].
The complexity of the Dynamic Algorithm in Stage II is $K \cdot \mathcal{O}(1)$, which is negligible compared with the complexity of the Dual Algorithm.

Finally, substituting the obtained subscription fees and price assignment, we can find the critical type $m^\ast$ through exhaustive search.
Obviously, we only need to search $K+1$ times for $m^\ast$.
In summary, the overall complexity of our proposed algorithm is $$ (K+1) \cdot K \cdot \mathcal{O}\left( \frac{1}{\epsilon^2} \right) .$$

\newpage

\section{Heterogeneous Mobility Scenario}\label{sec:hetemobility}

In the previous discussion, we consider the homogenous mobility scenario, where different APOs have the same mobility pattern. 
In this section, we will study a more general scenario with the heterogeneous mobility, where different APOs may have different mobility patterns, i.e., different probabilities of staying at home (or traveling outside).

Similar as the APO quality model, We consider a finite and discrete set $\mathcal{M}=\{1,2,\ldots,L\}$ of $L$ possible probabilities (of staying at home) for APOs, and denote $\eta_l, l\in \mathcal{M}$ as the $l$-th probability.
Without loss of generality, we assume:
$$
\eta_1 < \eta_2 <\cdots <\eta_L.
$$

Recall that in our previous analysis, all APOs are divided into $K$ types based on their provided Wi-Fi access qualities, and a type-$k$ APO's Wi-Fi access quality is denoted by $\theta_k, \forall k \in \K$.
In this section, each APO is characterized not only by his provided Wi-Fi access quality, but also by his mobility pattern.
Hence, all APOs can be divided into $K \times L$ types based on their provided Wi-Fi access qualities and their mobility patterns, where a type-$\{k,l\}$ APO provides the Wi-Fi access with quality $\theta_k$ and stays at home with probability $\eta_l$.
Note that $\theta_k$ and $\eta_l$ are private information of each APO, and are not known by the network operator.

\subsection{Contract Formulation}

To induce APOs to reveal their private information truthfully, the operator needs to design a contract item for each type of APO (characterized by $\theta_k$ and $\eta_l$).
Hence, the contract that the operator offers is
\vspace{-1mm}
\begin{equation}
\Phi = \{\phi_{k,l} \eq (p_{k,l},\delta_{k,l}):k \in \mathcal{K},l\in\mathcal{M}\}.
\vspace{-1mm}
\end{equation}
Here the contract item $\phi_{k,l}$ is designed for the type-$\{k, l\}$ APO.
A special combination $\phi_0 = (0,0)$ indicates the membership choice of Linus.
We let $N_{k,l}$ denote the number of the type-$\{k, l\}$ APOs.
We let $\LL(\Phi)$ denote the set of all Linus, and let $\BB(\Phi)$ denote the set of all Bills.
We further denote $\LL_l(\Phi)$ as the set of Linus with $\eta_l$, and denote $\BB_l(\Phi)$ as the set of Bills with $\eta_l$, for all $l \in \mathcal{M}$.
When there is no confusion, we will also write $\LL$ for $\LL(\Phi)$, and similarly for $\BB, \LL_l, \BB_l$.

Similar as in Section \ref{sec:contract}, we characterize the network operator's profit and each APO's payoff under a given feasible contract, where each APO chooses the contract item designed for his type.

The network operator's profit includes the profit achieved from all Linus' APs and the profit achieved from all Bill APs, and can be computed as follows:
\vspace{-2mm}
\begin{align}\label{eq:profit_new}
&\left[ \sum_{l=1}^L \sum_{k\in\LL_l} N_{k,l} \left( \sum_{j=1}^L  \sum_{i\in\BB_j} \frac{1-\eta_j}{N} N_{i,j} +a\right) p_0d_k(p_0)\right] \notag \\
&+\sum_{l=1}^L\sum_{k\in\mathcal{B}_l}N_{k,l}\delta_{k,l} .
\vspace{-2mm}
\end{align}

If a type-$\{k, l\}$ APO chooses to be a Linus, his payoff is the difference between utility and cost,
\vspace{-1mm}
\begin{equation}
\label{eq:profitLinusnew}
u_{k,l}(\phi_0) = \eta_l \Uhome + (1-\eta_l) \Uroam - \Cserve.
\vspace{-1mm}
\end{equation}

If a type-$\{k, l\}$ APO chooses to be a Bill, his payoff is the difference between his revenue, utility, and cost,
\vspace{-1mm}
\begin{equation}\label{eq:profitBillnew}
\begin{aligned}
u_{k,l}(\phi_{k,l};\Phi) = & \omega_l(\Phi) g_k(p_{k,l})-\delta_{k,l}-\beta_{k,l}(\Phi)
\\
& + \eta_l  \Uhome + (1-\eta_l)  \Uroam - \Cserve .
\end{aligned}
\vspace{-1mm}
\end{equation}
Here $\omega_l(\Phi)$ is the expected number of paying users (including others type-$\{k,l\}$ Bills characterized by $\theta_k$ and $\eta_l$, all other types of Bills, and all Aliens) accessing this AP:
\begin{equation}\label{eq:alphanew}
\displaystyle \omega_l(\Phi) \eq  \textstyle  \sum_{j=1}^L\sum_{i\in\mathcal{B}_j}\frac{1-\eta_j}{N}  N_{i,j} +\frac{1}{N}N_A - \frac{1-\eta_l}{N},
\end{equation}
and $\beta_{k,l}(\Phi)$ is the expected payment of this type-$\{k,l\}$ APO for accessing other APs:
\vspace{-1mm}
\begin{align}\label{eq:betaknew}
\beta_{k,l}(\Phi) \eq & \textstyle \frac{1-\eta_l}{N} \Big[ \sum_{j=1}^L\sum_{i\in\mathcal{B}_j} N_{i,j}g_i(p_{i,j}) \notag \\
& + \sum_{j=1}^L\sum_{i\in\mathcal{L}_j}N_{i,j}g_i(p_0) -g_k(p_{k,l}) \Big].
\vspace{-1mm}
\end{align}

For notational convenience, we denote
\vspace{-1mm}
\begin{align*}
& \mu(\Phi) =\sum_{j=1}^L \sum_{i\in\BB_j}\frac{1-\eta_j}{N}N_{i,j}+\frac{N_A}{N} , \\
& \nu(\Phi) =\sum_{j=1}^L \sum_{i\in\BB_j}N_{i,j}g_i(p_{i,j})+\sum_{j=1}^L \sum_{i\in\LL_j}N_{i,j}g_i(p_0) .
\vspace{-1mm}
\end{align*}
Then, a Bill APO's payoff can be written as
\begin{equation}\label{eq:profitBillnew2}
u_{k,l}(\phi_{k,l};\Phi)=\mu(\Phi) g_k(p_{k,l})-\delta_{k,l}-\frac{1-\eta_l}{N} \nu(\Phi).
\end{equation}
When there is no confusion, we will also write $\omega_l$ for $\omega_l(\Phi)$, and similarly for $\beta_{k,l}, \mu, \nu$.

\subsection{Feasibility of   Contract}

We now characterize the necessary conditions for a feasible contract.

\begin{lemma}\label{lemma_p_delta_new}
If a contract $\Phi$ is feasible, then
\vspace{-1mm}
$$
p_{k,l} > p_{i,j} \Longleftrightarrow \delta_{k,l} > \delta_{i,j},\quad \forall k \in \BB_l, i \in \BB_j, l,j\in\mathcal{M}.
\vspace{-1mm}
$$
\end{lemma}

\begin{lemma}\label{lemma_p_theta_new}
If a contract $\Phi$ is feasible, then
\begin{align*}
& \theta_k>\theta_i \Longrightarrow p_{k,l}>p_{i,l}, \forall k,i\in\BB_l, \mbox{ for the same } l \in \mathcal{M},  \\
& \eta_l > \eta_j \Longrightarrow p_{k,l} \geq  p_{k,j}, \forall l,j \in \mathcal{M}, \mbox{ for the same } k\in\BB_l, \BB_j, \\
& \eta_l < \eta_j \mbox{ and } \theta_k>\theta_i \Longrightarrow p_{k,l}>p_{i,j}, \forall k\in\BB_l, i\in\BB_j, l,j\in \mathcal{M}.
\end{align*}
\end{lemma}

Lemma \ref{lemma_p_delta_new} shows that in the heterogeneous mobility scenario, a larger Wi-Fi access price corresponds to a larger subscription fee (for Bill). 
Lemma \ref{lemma_p_theta_new} shows that with the same probability of staying at home, an APO who provides a higher Wi-Fi quality will be designed with a higher Wi-Fi access price.
With the same Wi-Fi quality, an APO who has a larger probability of staying at home will be designed with a higher Wi-Fi access price.\footnote{For an APO with a larger probability of staying at home, the average number of users accessing his AP is higher than the average number of users accessing an other APO (who has a relatively lower probability of staying at home).
Therefore, the operator can potentially set a higher Wi-Fi access price, due to the higher demand.}
What is most significant is the their relationship in Lemma \ref{lemma_p_theta_new}:
\emph{if an APO provides a higher Wi-Fi quality, even if he stays at home with a smaller probability, he will be designed with a higher Wi-Fi access price, which implies that the Wi-Fi quality is more important than the mobility pattern.}

\begin{lemma}\label{lemma_kc_new}
If a contract $\Phi$ is feasible, then there exists a critical APO type $m_l\in\{1,2,\ldots,K+1\}$ for each $\eta_l ~(\forall l\in\mathcal{M})$, such that $k \in \mathcal{L}_l$ for all $k < m_l$ and $k \in \mathcal{B}_l$ for all $k \geq m_l$, i.e.,
\begin{equation*}
\LL_l=\{1,2,\ldots,m_l-1\},\BB_l=\{m_l,m_l+1,\ldots,K\}, \forall l\in\mathcal{M}.
\end{equation*}
\end{lemma}

\begin{lemma}\label{lemma_m}
If a contract $\Phi$ is feasible, then
$$
m_l \leq m_j, \forall~ l>j \mbox{ where } l,j\in\mathcal{M}.
$$
\end{lemma}

Lemma \ref{lemma_kc_new} shows that for APOs with the same probability of staying at home, there exists a critical APO type: all APOs with types lower than the critical type will choose to be Linus, and all APOs with types higher than or equal to the critical type will choose to be Bills.
Lemma \ref{lemma_m} shows that APOs who stay at home with a larger probability  are more likely to choose to be Bills.

By Lemmas \ref{lemma_p_theta_new}, \ref{lemma_kc_new}, and \ref{lemma_m}, we can provide a lower bound and an upper bound for each Bill's price:
\vspace{-1mm}
\begin{equation}\label{eq:price_new}
\begin{cases}
& p_{k-1,L} \leq p_{k,1} \leq p_{k,2}, \forall k\in\BB_1, \\
& \max \left\{ p_{k,l-1},p_{k-1,L} \right\} \leq p_{k,l} \leq p_{k,l+1}, \\
& \quad \quad \quad \quad \forall k\in\BB_l, \forall 1<l<L, \\
& \max \left\{ p_{k,L-1},p_{k-1,L}  \right\} \leq p_{k,L}  \leq p_{k+1,j},\\
& \quad \quad \quad \quad \forall k\in\BB_L, j=\min_{i\in\mathcal{M}}\{i:k+1 \geq m_i\}.
\end{cases}
\vspace{-1mm}
\end{equation}
Recall that a Linus APO will choose the contract item $\phi_0=(0,0)$, i.e., set his price to be $p=0$.
Hence, if the type-$\{k-1,L\}$ APO chooses to be a Linus, then $p_{k-1,L}=0$.
Note that $p_{k-1,L} \leq p_{k,l}$ and $p_{k,L} \leq p_{k+1,j}$ are due to the third relationship in Lemma \ref{lemma_p_theta_new}, and $j=\min_{i\in\mathcal{M}}\{i:k+1 \geq m_i\}$ enables that the type-$\{k+1,j\}$ APO chooses to be a Bill and has the smallest probability of staying at home among Bills with $\theta_{k+1}$.
Intuitively, the relationships in \eqref{eq:price_new} can be interpreted from two aspects: (a) a Bill APO who provides a higher Wi-Fi quality will be designed with a higher Wi-Fi access price, regardless of the mobility pattern; (b) with the same Wi-Fi access quality, a Bill APO who has a larger probability of staying at home will be designed with a higher Wi-Fi access price.

To illustrate the relationship among all APOs' prices, we provide a simple example where $K=4, L=5, m_1=5, m_2=4, m_2=3$, and $m_4=m_5=1$.
The price matrix of all APOs in the toy example is as follows:
\vspace{-1mm}
\begin{equation*}
\left(
  \begin{array}{ccccc}
    0 & 0 & 0 & p_{1,4} & p_{1,5}\\
    0 & 0 & 0 & p_{2,4} & p_{2,5}\\
    0 & 0 & p_{3,3} & p_{3,4} & p_{3,5}\\
    0 & p_{4,2} & p_{4,3} & p_{4,4} & p_{4,5}
  \end{array}
\right)
\vspace{-1mm}
\end{equation*}
where APOs in the same column have the same $\eta_l ~(l\in\M)$, and APOs in the same row have the same $\theta_k ~(k \in \K)$.
The $0$ elements represent the choices of Linus, while the elements $p_{k,l} ~(\forall k \in \K, l\in\M)$ represent the choices of Bill.
From \eqref{eq:price_new}, we know that Bills' prices satisfy:
\vspace{-1mm}
\[ p_{1,4} \leq p_{1,5} \leq p_{2,4} \leq p_{2,5} \leq p_{3,3} \leq \cdots \leq p_{4,5}.
\vspace{-1mm} \]

We now characterize the sufficient and necessary conditions for a feasible contract.
\begin{theorem}[Feasible Contract]\label{lemma_Feasibility_new}
A contract $\Phi = \{\phi_{k,l}:k \in \mathcal{K},l\in\mathcal{M}\}$ is feasible, if and only if the following conditions hold:
\begin{align}
& \mathcal{L}_l=\{1,\ldots,m_l-1\}, \mathcal{B}_l=\{m_l ,\ldots,K\},\ \notag \\
& \quad \quad ~~  m_l\in \{1,...,K+1\},\forall l\in\mathcal{M}, \label{eq:con1new} \\
& m_1 \geq m_2 \geq m_3 \geq \cdots \geq m_L,\  \label{eq:con11} \\
& 0 \leq p_0 \leq p_{\max},0 \leq p_{k,l} \leq p_{\max},~ \forall k\in\mathcal{K}, \forall l\in\mathcal{M},\label{eq:con2new}\\
& p_{k-1,L} \leq p_{k,1} \leq p_{k,2} ,~ \forall k\in\BB_1, \label{eq:con2new2} \\
& \max \left\{ p_{k,l-1},p_{k-1,L} \right\} \leq p_{k,l} \leq p_{k,l+1} , \notag \\
& \quad \quad \quad~ \forall k\in\BB_l, \forall 1<l<L, \label{eq:con2new3} \\
& \max \left\{ p_{k,L-1},p_{k-1,L}  \right\} \leq p_{k,L}  \leq p_{k+1,j} ,\notag \\
& \quad \quad \quad~\forall k\in\BB_L, j=\min_{i\in\mathcal{M}}\{i:k+1\geq m_i\},  \label{eq:con2new4} \\
& \delta_{k,l} \geq \mu(\Phi) g_i(p_{k,l})-\frac{1-\eta_j}{N}\nu(\Phi),\ \notag\\
& \quad \quad \quad~\forall k\in\BB_l, i\in\LL_j, l,j\in \mathcal{M},\label{eq:con3new} \\
& \delta_{k,l} \leq \omega_l(\Phi) g_k(p_{k,l})-\beta_{k,l}(\Phi),\forall k\in\BB_l, l\in \mathcal{M}, \label{eq:con4new} \\
& \omega_j(\Phi) \left( g_i(p_{k,l})-g_i(p_{i,j}) \right) \leq \delta_{k,l}-\delta_{i,j} \notag \\
&~ \leq \omega_l(\Phi) \left( g_k(p_{k,l})-g_k(p_{i,j}) \right),  \forall k\in\BB_l, i\in\BB_j, l,j\in \mathcal{M}. \label{eq:con5new}
\end{align}
\end{theorem}


\subsection{Optimal Contract Design}

The optimal contract is the one that maximizes the network operator's profit.
Given the network operator's profit \eqref{eq:profit_new} and using Theorem \ref{lemma_Feasibility_new}, we can formulate the following contract optimization problem:
\begin{problem}[\textbf{Optimal Contract (Heterogeneous Mobility)}]
\begin{align}
& \displaystyle \max ~ \left[ \sum_{l=1}^L \sum_{k\in\LL_l} N_{k,l} \left( \sum_{j=1}^L  \sum_{i\in\BB_j} \frac{1-\eta_j}{N} N_{i,j}+a\right) p_0d_k(p_0)\right] \notag \\
& \quad \quad ~~ +\sum_{l=1}^L\sum_{k\in\mathcal{B}_l}N_{k,l}\delta_{k,l} \notag\\
& ~\mbox{s.t.} ~~~ \eqref{eq:con1new}-\eqref{eq:con5new} \notag \\
& ~\mbox{var:} ~~~ \{m_l:\forall l \in \mathcal{M}\}, p_0,\ \{(p_{k,l},\delta_{k,l}):\forall l \in \mathcal{M},k\in\BB_l \}. \notag
\end{align}
\end{problem}

Next we define the feasible price assignment under heterogeneous mobility, which will be used in later discussions.

\begin{definition}[Feasible Price Assignment in Heterogeneous Mobility]\label{def:fp_new}
A price assignment $\{p_0,$ $p_{k,l}:l \in \mathcal{M},k\in\BB_l \}$ is \emph{feasible} if it satisfies constraints \eqref{eq:con2new}, \eqref{eq:con2new2}, \eqref{eq:con2new3}, and \eqref{eq:con2new4}.
\end{definition}

We solve Problem 3 in the following steps similar as in Section \ref{sec:OptimalContract}.
First, we derive a closed-form solution of the best subscription fees $\{\delta_{k,l}:l \in \mathcal{M},k\in\BB_l\}$, given the feasible price assignment $\{p_0,p_{k,l}:l \in \mathcal{M},k\in\BB_l \}$ and critical types $\{m_l:l \in \mathcal{M}\}$.
Then, we propose an algorithm to solve the optimal price assignment $\{p_0,p_{k,l}:l \in \mathcal{M},k\in\BB_l \}$, given the critical types $\{m_l:l \in \mathcal{M}\}$.
Finally, based on the best subscription fees and price assignment, we search the critical types $\{m_l:l \in \mathcal{M}\}$ for the optimal contract.

\begin{figure*}[t]
\vspace{-4mm}
\begin{align}
&f_{m_j,j}(p_{m_j,j})=\textstyle \left( \frac{1-\eta_j}{N}+\omega_j \right) s_{m_j,j}   g_{m_j}(p_{m_j,j}) -\omega_{j+1} s_{m_j,j+1} g_{m_j}(p_{m_j,j}),\  j=\min_{i\in\mathcal{M}}\{i:m_i=m_L\}, \label{eq:f_new1}
\\
&f_{k,l}(p_{k,l})=  \omega_l s_{k,l} g_k(p_{k,l})- \omega_{l+1}s_{k,l+1}g_{k}(p_{k,l}) , \ l<L,k\in\BB_l , \label{eq:f_new2}
\\
&f_{k,L}(p_{k,L})=  \omega_L s_{k,L} g_k(p_{k,L})- \omega_{i}s_{k+1,i}g_{k+1}(p_{k,L}) , \ k \in\BB_L , i= \min_{l\in\mathcal{M}}\{l:k+1 \geq m_l\}, \label{eq:f_new3}
\\
&f_{K,L}(p_{K,L})= \omega_L N_{K,L} g_K(p_{K,L}). \label{eq:f_new4}
\end{align}
\hrulefill
\vspace{-4mm}
\end{figure*}

We first derive the best subscription fees, given the feasible price assignment and critical types.
\begin{problem}[\textbf{Optimal Subscription Fees (Heterogeneous Mobility)}]
\begin{align}
& \displaystyle \max ~~ \sum_{l=1}^L\sum_{k\in\mathcal{B}_l}N_{k,l}\delta_{k,l} \notag\\
& ~\mbox{s.t.} ~~~~ \eqref{eq:con3new},\eqref{eq:con4new},\eqref{eq:con5new} \notag \\
& ~\mbox{var:} ~~~ \delta_{k,l}, \forall l \in \mathcal{M},k\in\BB_l . \notag
\end{align}
\end{problem}

Lemma \ref{lemma_optimal_delta_new} gives the optimal solution of Problem 5.

\begin{lemma}
\label{lemma_optimal_delta_new}
Given critical types $\{m_l:\forall l \in \mathcal{M}\}$ and a feasible price assignment $\{p_0,p_{k,l}:\forall l \in \mathcal{M},k\in\BB_l\}$,
the optimal subscription fees $\{{\delta}_{k,l}^\ast:\forall l \in \mathcal{M},k\in\BB_l\}$ are given by:
\begin{align}
& {\delta}^\ast_{m_j,j} =\omega_j g_{m_j}(p_{m_j,j})-\beta_{m_j,j}, j=\min_{i\in\mathcal{M}}\{i:m_i=m_L\}, \label{eq:delta_new1} \\
& {\delta}^\ast_{k,l} = {\delta}^\ast_{k,l-1}+\omega_l \left( g_k(p_{k,l})-g_k(p_{k,l-1}) \right), \notag \\
& \quad \quad \quad \forall l\in\mathcal{M},k \geq m_{l-1},  \label{eq:delta_new2} \\
& {\delta}^\ast_{k,l}=\delta^\ast_{k-1,L}+\omega_l \left( g_k(p_{k,l})-g_{k-1}(p_{k-1,L}) \right) , \notag \\
& \quad \quad \quad \forall l\in\mathcal{M},k < m_{l-1}.\label{eq:delta_new3}
\end{align}
\end{lemma}

Note that $\omega_j,\omega_l$ and $\beta_{m_j,j}$ depend on $\{m_l:\forall l \in \mathcal{M}\}$ and $\{p_0,p_{k,l}:\forall l \in \mathcal{M},k\in\BB_l\}$, and can be calculated by \eqref{eq:alphanew} and \eqref{eq:betaknew}.
Define $s_{k,l}= \sum_{j\geq l}N_{k,j}+\sum_{j\in\mathcal{M}}\sum_{i\in\BB_j,i>k}N_{i,j} $ for all $l\in\mathcal{M},k\in\BB_l$.
Based on Lemma \ref{lemma_optimal_delta_new}, we can rewrite the network operator's profit from Bills as follows:
\[
\sum_{l=1}^L\sum_{k\in\mathcal{B}_l}N_{k,l}\delta_{k,l}^\ast \eq \sum_{l=1}^L\sum_{k\in\mathcal{B}_l}f_{k,l}(p_{k,l}),
\]
where $f_{k,l}(p_{k,l})$ only depends on $p_{k,l}$, and is defined in \eqref{eq:f_new1}, \eqref{eq:f_new2}, \eqref{eq:f_new3}, and \eqref{eq:f_new4}. 

Next we solve the operator's optimal prices, given critical types $\{m_l:\forall l \in \mathcal{M}\}$.
Substitute the best subscription fees $\{\delta_{k,l}^\ast:\forall l \in \mathcal{M},k\in\BB_l \}$ in Lemma \ref{lemma_optimal_delta_new}, we have the following optimization problem.
\begin{problem}[\textbf{Optimal Price Assignment (Heterogeneous Mobility)}]
\begin{align}
& \displaystyle \max ~~ \left[ \sum_{l=1}^L \sum_{k\in\LL_l} N_{k,l} \left( \sum_{j=1}^L  \sum_{i\in\BB_j} \frac{1-\eta_j}{N} N_{i,j}+a\right) p_0d_k(p_0)\right] \notag \\
& \quad \quad ~~+\sum_{l=1}^L\sum_{k\in\mathcal{B}_l} f_{k,l}(p_{k,l}) \notag\\
& ~\mbox{s.t.} ~~~~  \eqref{eq:con2new},\eqref{eq:con2new2},\eqref{eq:con2new3},\eqref{eq:con2new4} \notag \\
& ~\mbox{var:} ~~~ p_0, \ \{p_{k,l}:\forall l \in \mathcal{M},k\in\BB_l \}. \notag
\end{align}
\end{problem}


Finally, after obtaining the subscription fees and the price assignment, we can sequentially search for the optimal critical type $\{m_l^\ast:l \in \mathcal{M}\}$.
Due to constraints \eqref{eq:con1new} and \eqref{eq:con11}, we only need to search ${(K+1)(K+2)}/{2}$ times for $\{m_l^\ast:l \in \mathcal{M}\}$.

\subsection{More Simulations under Heterogeneous Mobility}

\begin{figure}[t]
  \centering
\includegraphics[width=0.35\textwidth]{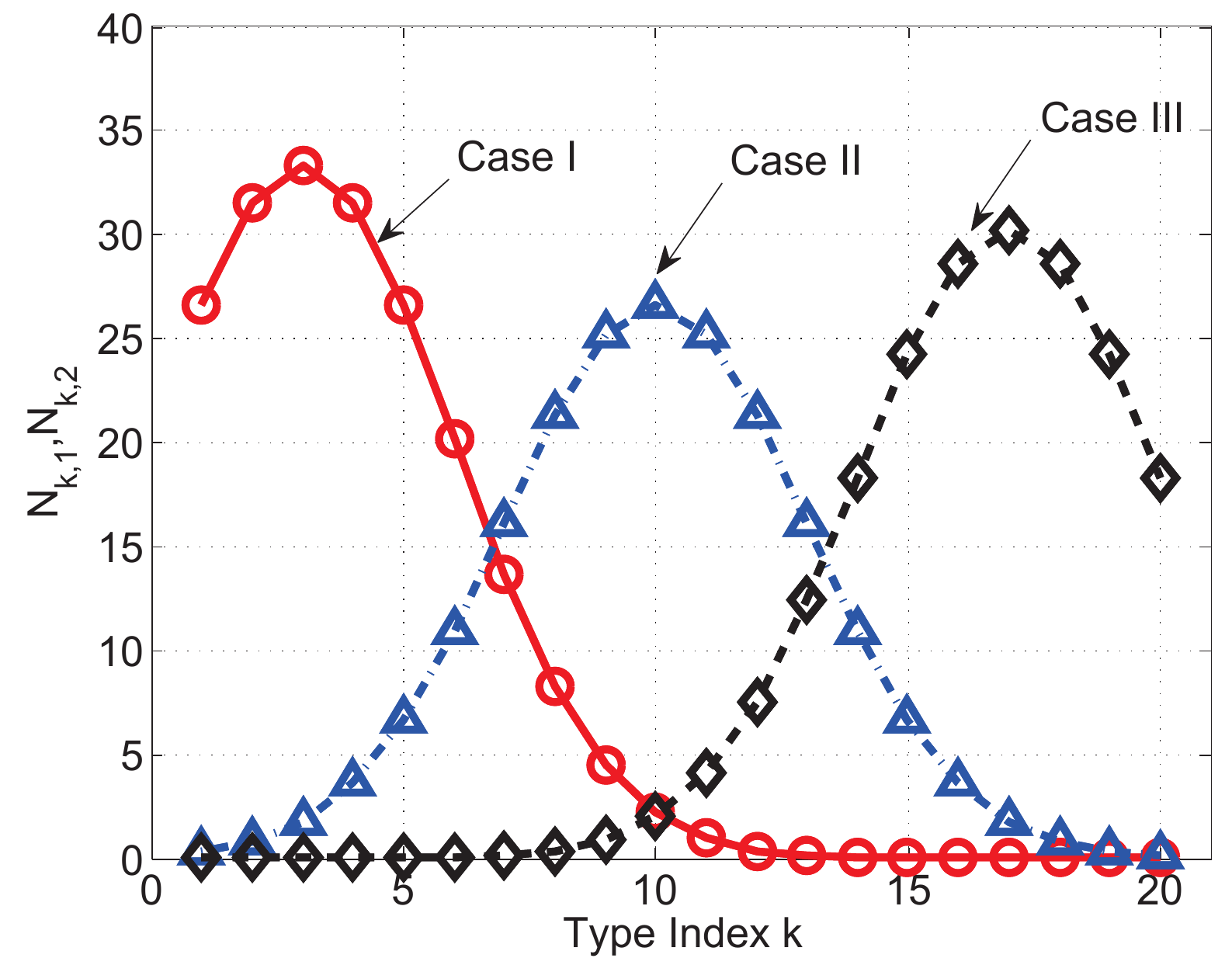}
  \caption{Distribution of APOs (Case I: low type APO dominant; Case II: medium type APO dominant; Case III: high type APO dominant)}\label{fig:TwoDimenAPD}
\end{figure}

\begin{figure*}[t]
\centering
\begin{minipage}[t]{0.3 \linewidth}
\centering
\includegraphics[width=1\textwidth]{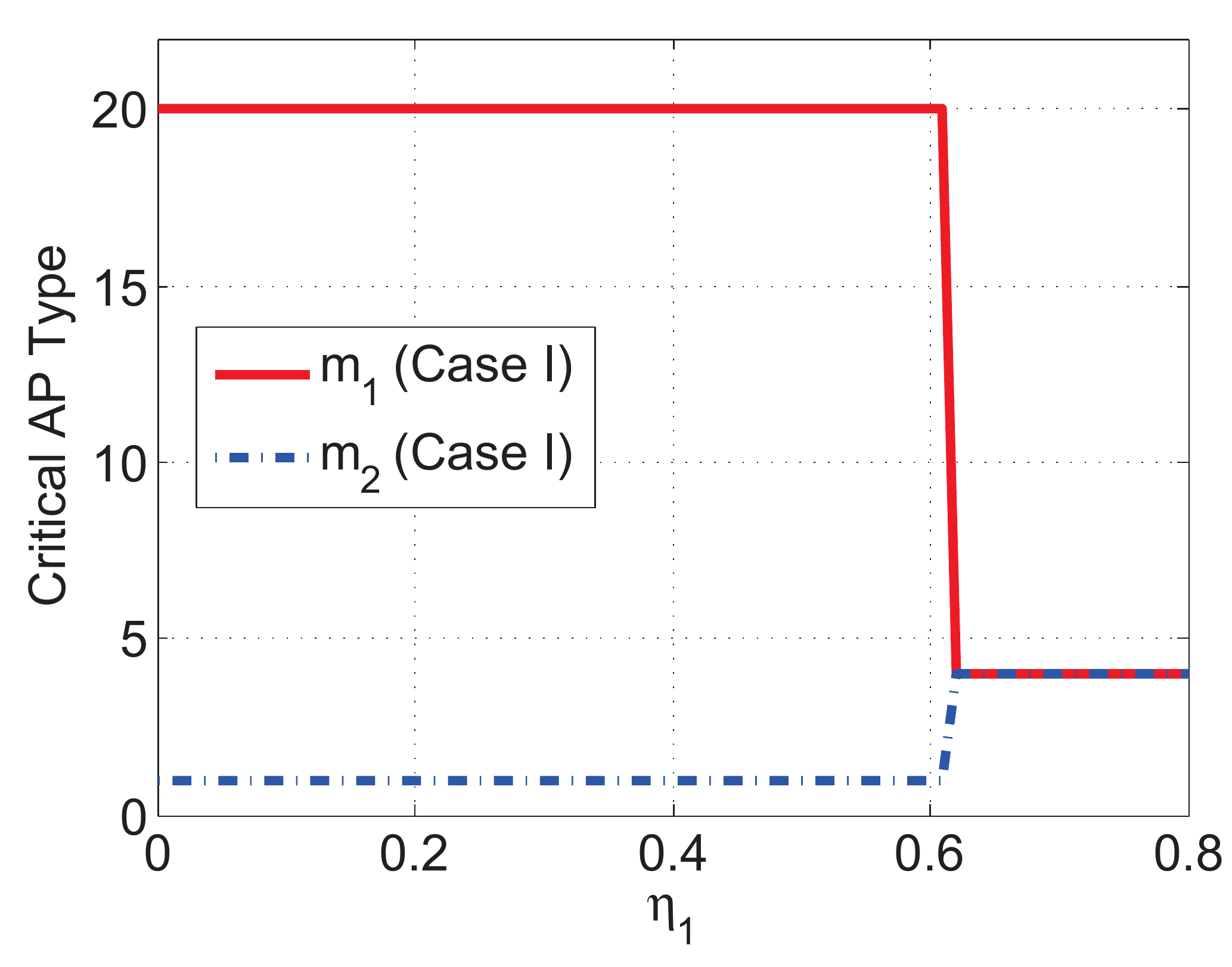}
  \caption{Impact of $\eta_1$ to Critical AP Types (Case I: low type APO dominant)}\label{fig:eta1m1m2a}
\end{minipage}
\begin{minipage}[t]{0.03 \linewidth}
~
\end{minipage}
\begin{minipage}[t]{0.3 \linewidth}
\centering
\includegraphics[width=1\textwidth]{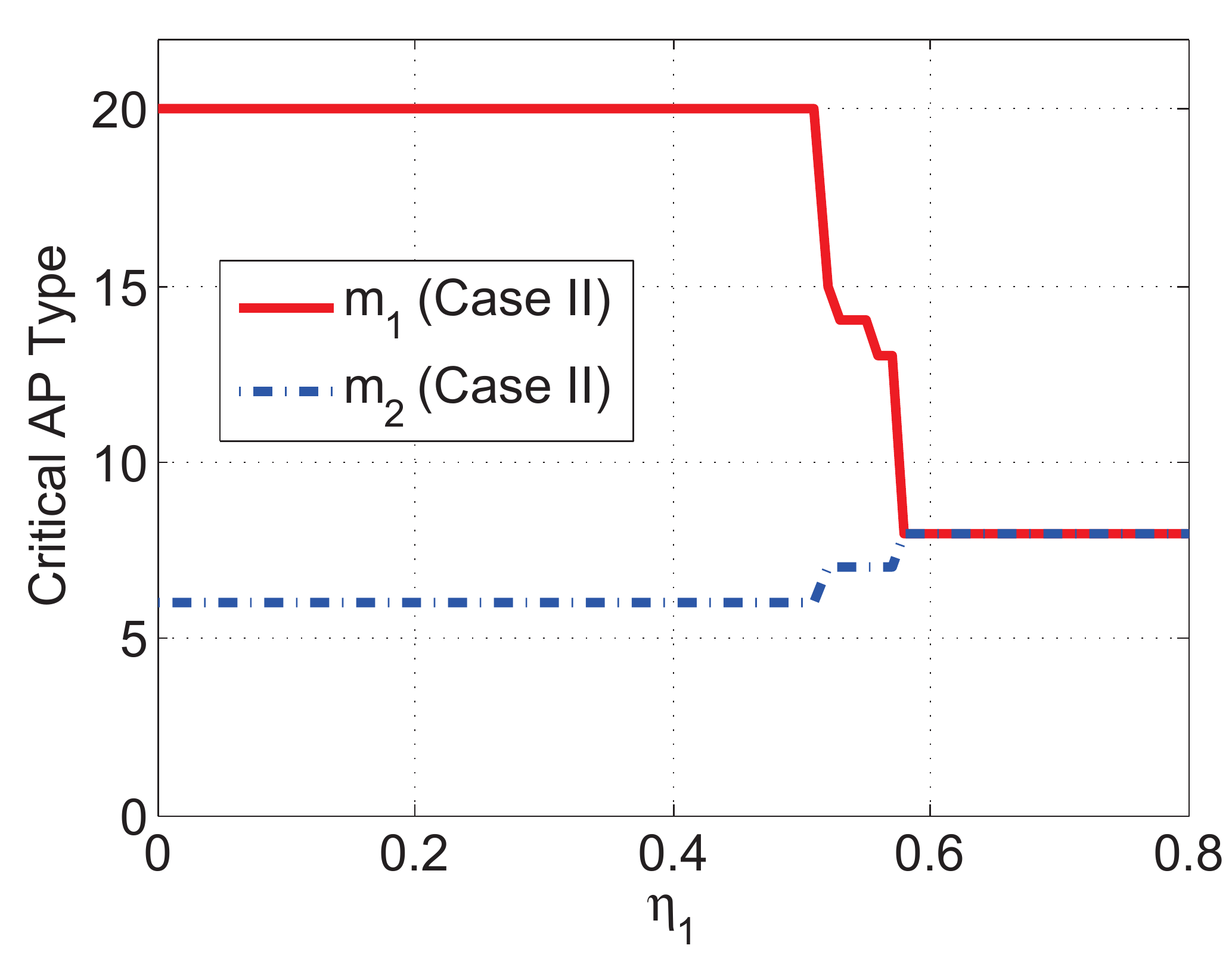}
  \caption{Impact of $\eta_1$ to Critical AP Types (Case II: medium type APO dominant)}\label{fig:eta1m1m2b}
\end{minipage}
\begin{minipage}[t]{0.03 \linewidth}
~
\end{minipage}
\begin{minipage}[t]{0.3 \linewidth}
\centering
\includegraphics[width=1\textwidth]{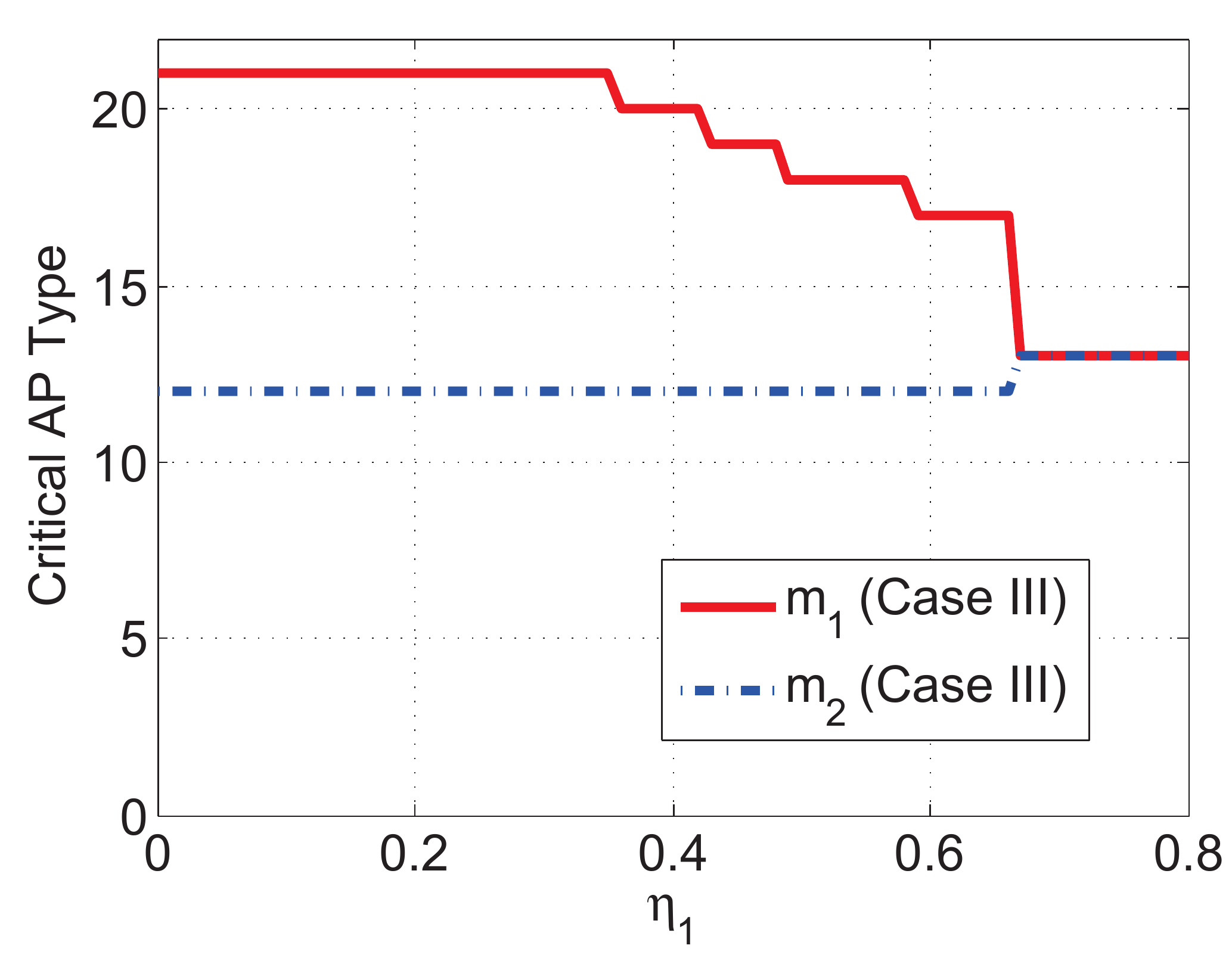}
  \caption{Impact of $\eta_1$ to Critical AP Types (Case III: high type APO dominant)}\label{fig:eta1m1m2c}
\end{minipage}
\end{figure*}

In Section \ref{sec:simu_hete}, we have shown how the critical AP types $m_1$ and $m_2$ change with the mobility parameters $\boldsymbol{\eta}$ under the uniform APO type distribution, where the number of each type of APOs is the same $N_{k,l}=100, \forall k\in\K, l\in\M$. 
In this section, we provide more simulation results under three different APO type distributions illustrated in Figure \ref{fig:TwoDimenAPD}. 
We show the new simulation results in Figures \ref{fig:eta1m1m2a}-\ref{fig:eta1m1m2c}, and can obtain the following observations.

First, the critical AP type $m_1$ decreases with $\eta_1$, i.e., more APOs who stay at home with a probability $\eta_1$ choose to be Bills as $\eta_1$ increases.
The reason is that as $\eta_1$ increases, APOs with $\eta_1$ stay at home with a larger probability and travel to other APs with a smaller probability.
So APOs with $\eta_1$ prefer to be Bills (to gain profit at her own APs) rather than be Linus (to enjoy free Wi-Fi at other APs).
Hence, more APOs with $\eta_1$ choose to be Bills as $\eta_1$ increases.

Second, the critical AP type $m_2$ increases with $\eta_1$, i.e., less APOs who stay at home with probability a $\eta_2$ choose to be Bills as $\eta_1$ increases.
The reason is that as $\eta_1$ increases from a medium to a large value, Bill APOs with $\eta_1$ stay at home with a larger probability, so the profit achieved by Bills from these paying users (Bill APOs with $\eta_1$) will decrease.
Hence, less APOs with $\eta_2$ choose to be Bills.

Third, more APOs will choose to be Linus in Case III (high type APO dominant) than in the other two cases.
The reason is that higher type APOs charge higher prices (and many of them charge $p_{\max}$).
In the high type APO dominant case (Case III), on average it is more expensive for a Bill to access other APs.
As a result, a lower type APO's payoff as a Bill is smaller due to the higher payment on other APs.
Hence, more APOs will choose to be Linus.

\newpage

\section{Proofs}

\subsection{Proof for Lemma 1}

\begin{proof}
We prove this lemma by using the IC constraint.

First, we show that if $p_i>p_j$, then $\delta_i>\delta_j$.
For any type-$j$ Bill (APO), the following IC constraint must be satisfied:
\[ u_j(p_j,\delta_j; \Phi) \geq u_j(p_i,\delta_i; \Phi),\quad \forall i \in \mathcal{B}. \]
That is,
$\omega(\Phi) g_j(p_j)-\delta_j-\beta_j(\Phi) \geq \omega(\Phi)  g_j(p_i)-\delta_i -\beta_j(\Phi),
$
which implies that
$ \delta_i - \delta_j \geq \omega(\Phi) ( g_j(p_i)- g_j(p_j) ) > 0. $

Then, we show that if $\delta_i > \delta_j$, then $p_i > p_j$.
For any type-$i$ Bill (APO), the following IC constraint must be satisfied:
\[ u_i(p_i,\delta_i; \Phi) \geq u_i(p_j,\delta_j; \Phi), \quad \forall j \in \mathcal{B}. \]
That is,
$\omega(\Phi) g_i(p_i)-\delta_i-\beta_i(\Phi) \geq \omega(\Phi)  g_i(p_j)-\delta_j -\beta_i(\Phi),$
which  implies that
$ g_i(p_i)-g_i(p_j) \geq  \frac{\delta_i - \delta_j}{\omega(\Phi)} > 0.$
According to Assumption 1.(b), we have
$ p_i > p_j . $
\end{proof}

\subsection{Proof of Lemma 2}

\begin{proof}
We prove the lemma by the contradiction principle.
Assume to the contrary that there exists $\theta_i > \theta_j$ and $p_i < p_j$.

From the IC constraints for the type-$i$ ($\forall i \in \mathcal{B}$) APOs and the type-$j$ ($\forall j \in \mathcal{B}$) APOs, we have
\[u_i(\phi_i; \Phi) \geq u_i(\phi_j; \Phi),~u_j(\phi_j; \Phi) \geq u_j(\phi_i; \Phi),\]
which are equivalent to
\begin{align}
& \omega(\Phi) g_i(p_i) - \delta_i -\beta_i(\Phi) \geq \omega(\Phi) g_i(p_j) -\delta_j-\beta_i(\Phi), \notag\\
& \omega(\Phi) g_j(p_j) - \delta_j -\beta_j(\Phi) \geq \omega(\Phi) g_j(p_i) - \delta_i-\beta_j(\Phi). \notag
\end{align}
Combining the above two inequations, we have
\[ g_i(p_i)-g_i(p_j) \geq g_j(p_i)-g_j(p_j). \]
This contracts with Assumption 1(c).
Hence we complete the proof.
\end{proof}

\subsection{Proof of Lemma 3}

\begin{proof}
We prove the lemma by the contradiction principle.
Suppose a type-$j$ APO chooses to be Linus, but a type-$i$ ($i<j$) APO chooses to be Bill.

From the IC constraint for the type-$j$ (Linus) APO, we have
\[ u_j(p_i,\delta_i; \Phi) \leq u_j(0,0),\]
equivalently,
\[ \mu(\Phi) g_j(p_i)-\delta_i-\nu(\Phi)\leq 0 .\]

Then for the type-$i$ ($i < j$) (Bill) APO, we have
\begin{align*}
u_i(p_i,\delta_i; \Phi) &= \mu(\Phi) g_i(p_i)-\delta_i-\nu(\Phi) \\
& < \mu(\Phi) g_j(p_i)-\delta_i-\nu(\Phi)  = u_j(p_i,\delta_i; \Phi) \leq 0
\end{align*}
which contradicts with the IR constraint for the type-$i$ (Bill) APO.

Therefore, for a feasible contract $\Phi=\{\phi_k:\forall k \in \mathcal{K}\}$, there exists a critical AP type $m$ such that $k \in \mathcal{L}$ for all $k < m$, and $k \in \mathcal{B}$ for all $k \geq m$.
\end{proof}

\subsection{Proof of Theorem 1}

\begin{proof}
We first prove that the conditions in Theorem 1 are \emph{sufficient} by mathematical induction.

We denote $\Phi(m-1+n)$ as a subset of $\Phi$ which contains the first $m-1+n$ contract items in $\Phi$, i.e., $\Phi(m-1+n)=\left\{(0,0):k\in \LL \right\} \cup \left\{(p_k,\delta_k):k\in \BB \right\}$ with $|L|=m-1$ and $|B|=n.$
Let $\Phi(m-1+n)$ be a contract for the network which contains the first $m-1+n$ types of AP owners of the original network.

Obviously, $m=K+1$ indicates that all AP owners choose to be Linus.
In the following proof, we consider the case where $m\leq K$.

We first verify that $\Phi(m)=\{(0,0):\forall k=1,2,\ldots,m-1\}\cup\{(p_{m},\delta_{m})\}$ is feasible, where only the type-$m$ APOs choose to be Bills, while other types of APOs choose to be Linus.
The conditions for such a contract to be feasible are the IR and IC constraints for all APOs.
Obviously, conditions in Theorem 1 imply
\begin{align}
& \omega(\Phi) g_{m}(p_{m})-\delta_{m}-\beta_m(\Phi) \geq 0, \notag \\
& \mu(\Phi) g_{j}(p_{m})-\delta_{m}-\nu(\Phi) \leq 0,~\forall j \in \mathcal{L}. \notag
\end{align}
Thus the IR and IC constraints for all APOs are satisfied and $\Phi(m)$ is a feasible contract.

We then show that if $\Phi(m-1+k)=\{(0,0):\forall i=1,\ldots,m-1\}\cup\{(p_i,\delta_i):\forall i=m,\ldots,m-1+k\}~(k \geq 1)$ is a feasible contract, then $\Phi(m+k)$ is also feasible.
To achieve this, we need to prove that (I) for the new type $\theta_{m+k}$, the IC and IR constraints are satisfied, i.e.,
\[
\begin{cases}
& \omega(\Phi) g_{m+k}(p_{m+k})-\delta_{m+k} -\beta_{m+k}(\Phi)  \geq \omega(\Phi) g_{m+k}(p_i)\\
&~~-\delta_i -\beta_{m+k}(\Phi),\forall i=m, \ldots ,m-1+k,\\
& \omega(\Phi) g_{m+k}(p_{m+k})-\delta_{m+k} -\beta_{m+k}(\Phi) \geq 0,
\end{cases}
\]
and (II) for the existing types $\theta_1,\ldots ,\theta_{m-1},\theta_{m},\ldots,\theta_{m-1+k}$, the IC constraints are still satisfied in the presence of type $\theta_{m+k}$, i.e.,
\[
\begin{cases}
& \displaystyle \mu(\Phi) g_j(p_{m+k})-\delta_{m+k}-\nu(\Phi) \leq 0,~\forall j=1, \ldots ,m-1,\\
& \omega(\Phi) g_i(p_i)-\delta_i -\beta_i(\Phi) \geq \omega(\Phi) g_i(p_{m+k})-\delta_{m+k}\\
&~~-\beta_i(\Phi),\forall i=m,\ldots , m-1+k .
\end{cases}
\]

Next, we prove (I) and (II) separately.

Proof of I:
Since
\begin{align*}
& \delta_{m+k}-\delta_i \leq \omega(\Phi) \left( g_{m+k}(p_{m+k})-g_{m+k}(p_i) \right), \\
& ~~\forall i=m,\ldots , m-1+k,
\end{align*}
we have
\begin{align*}
& \omega(\Phi) g_{m+k}(p_{m+k})-\delta_{m+k} -\beta_{m+k}(\Phi) \geq \omega(\Phi) g_{m+k}(p_i)\\
&~~-\delta_i -\beta_{m+k}(\Phi),~\forall i=m, \ldots ,m-1+k,
\end{align*}
which proves that the IC constraint is satisfied.
Since
\[\delta_{m+k} \leq \omega(\Phi) g_{m+k}(p_{m+k})-\beta_{m+k}(\Phi)\]
we have
\[ \omega(\Phi) g_{m+k}(p_{m+k})-\delta_{m+k} -\beta_{m+k}(\Phi)\geq 0, \]
which proves that the IR constraint is satisfied.

Proof of II:
Since
\[\delta_{m+k} \geq \mu(\Phi) g_j(p_{m+k})-\nu(\Phi),\forall j=1,\ldots,m-1,\]
we have
\[ \mu(\Phi) g_j(p_{m+k})-\delta_{m+k}-\nu(\Phi) \leq 0,~\forall j=1, \ldots ,m-1,\]
which proves that the IC constraints for the Linus AP owners in the set $\mathcal{L}=\{1,\ldots,m-1\}$ are satisfied.
Since $\delta_{m+k}-\delta_i \geq \omega(\Phi) \left( g_i(p_{m+k})-g_i(p_i) \right) ,~\forall i=m,\ldots , m-1+k$, we have
\begin{align*}
& \omega(\Phi) g_i(p_i)-\delta_i -\beta_i(\Phi) \geq \omega(\Phi) g_i(p_{m+k})-\delta_{m+k}-\beta_i(\Phi),\\
& \forall i=m,\ldots , m-1+k,
\end{align*}
which proves the IC constraints for the Bill AP owners in the set $\mathcal{B}=\{m,\ldots,m-1+k\}$.

Up to present, we have proved that (i) $\Phi(m)$ is feasible, and (ii) if $\Phi(m-1+k)$ (for $k\geq 1$) is feasible, then $\Phi(m+k)$ is feasible. It follows that $\Phi=\Phi(K)$ is feasible.

Now we prove that the conditions in Theorem 1 are \emph{necessary}.
Note that (13) has been proved by Lemma 3, and (14) has been proved by Lemma 2.
We now prove that (15), (16), and (17) are necessary by using the IC and IR constraints APOs in the network.
For any type-$j$ ($\forall j \in \mathcal{L}$) Linus APO, his payoff by choosing to be a Linus is no smaller than the payoff by choosing any contract item $(p_k,\delta_k) , ~\forall k \in \mathcal{B}$ designed for a Bill under a feasible contract, i.e., the following IC constraint must be satisfied:
\[ u_j(p_k,\delta_k; \Phi) \leq u_j(0,0),\]
which is
$$\mu(\Phi) g_j(p_k)-\delta_k-\nu(\Phi)\leq 0.$$
This implies that (15) in Theorem 1 is satisfied:
\[ \delta_k \geq \mu(\Phi) g_j(p_k)-\nu(\Phi),\ \forall k \in \mathcal{B}, \forall j \in \mathcal{L}. \]
For any type-$k$ ($\forall k \in \mathcal{B}$) Bill APO, the following IR constraint must be satisfied:
\[ u_k(p_k,\delta_k; \Phi) \geq 0, \]
which is
$$\omega(\Phi) g_k(p_k)-\delta_k-\beta_k(\Phi) \geq 0.$$
This implies that (16) in Theorem 1 is satisfied:
\[ \delta_k \leq \omega(\Phi) g_k(p_k)-\beta_k(\Phi),\ \forall k \in \mathcal{B}. \]
For any type-$k$ ($\forall k \in \mathcal{B}$) Bill APO and type-$i$ ($\forall i<k, i \in \mathcal{B}$) Bill APO, the following IC constraints must be satisfied:
\begin{align*}
& u_k(p_k,\delta_k; \Phi) \geq u_k(p_i,\delta_i; \Phi), \\
& u_i(p_i,\delta_i; \Phi) \geq u_i(p_k,\delta_k; \Phi),
\end{align*}
equivalently,
\begin{align*}
&  \omega(\Phi) g_k(p_k)-\delta_k-\beta_k(\Phi) \geq \omega(\Phi)  g_k(p_i)-\delta_i -\beta_k(\Phi), \\
&  \omega(\Phi) g_i(p_i)-\delta_i-\beta_i(\Phi) \geq \omega(\Phi)  g_i(p_k)-\delta_k -\beta_i(\Phi).
\end{align*}
This implies that (17) in Theorem 1 is satisfied:
\begin{align*}
&\omega(\Phi) \left( g_i(p_k)-g_i(p_i) \right) \leq \delta_k-\delta_i \leq \omega(\Phi)\left( g_k(p_k)-g_k(p_i) \right),\\
&~ \forall k,i \in \mathcal{B},i<k.
\end{align*}

\end{proof}

\subsection{Proof of Lemma 4}

\begin{proof}
We first prove that the subscription fees in \eqref{eq:delta_1} and \eqref{eq:delta_k} form a feasible subscription fee assignment, i.e., they satisfy the conditions \eqref{eq:con3}, \eqref{eq:con4}, and \eqref{eq:con5}.
Obviously, ${\delta}^*_{m}$ in \eqref{eq:delta_1} is feasible, that is, it satisfies \eqref{eq:con3}--\eqref{eq:con5}.
For ${\delta}^*_{m+1}={\delta}^*_{m}+\omega \left( g_{m+1}(p_{m+1})-g_{m+1}(p_{m}) \right])$, we first prove that it satisfies \eqref{eq:con3}.
We know that
\begin{align*}
& {\delta}^*_{m+1}-{\delta}^*_{m}=\omega \left( g_{m+1}(p_{m+1})-g_{m+1}(p_{m}) \right), \\
& \underline{\delta}_{m+1}-\underline{\delta}_{m}=\omega \left( g_{m-1}(p_{m+1})-g_{m-1}(p_{m}) \right).
\end{align*}
Given Assumption 1(c), the above two equations lead to:
\[{\delta}^*_{m+1} \geq \underline{\delta}_{m+1} + {\delta}^*_{m}-\underline{\delta}_{m} \geq \underline{\delta}_{m+1},\]
which proves that ${\delta}^*_{m+1}$ satisfies \eqref{eq:con3}.
We then prove that ${\delta}^*_{m+1}$ satisfies \eqref{eq:con4}:
\begin{align*}
{\delta}^*_{m+1} =&  \omega g_{m}(p_{m})-\beta_k + \omega \left( g_{m+1}(p_{m+1})-g_{m+1}(p_{m}) \right) \\
=& \mu g_{m}(p_{m})-\nu + \omega \left( g_{m+1}(p_{m+1})-g_{m+1}(p_{m}) \right) \\
\leq& \mu g_{m}(p_{m})-\nu + \mu \left( g_{m+1}(p_{m+1})-g_{m+1}(p_{m}) \right) \\
=& \mu g_{m+1}(p_{m+1}) - \mu \left( g_{m+1}(p_{m}) - g_{m}(p_{m}) \right) -\nu \\
\leq& \mu g_{m+1}(p_{m+1}) -\nu \\
=& \omega g_{m+1}(p_{m+1}) -\beta_{m+1}= \bar{\delta}_{m+1}.
\end{align*}
Finally, ${\delta}^*_{m+1}$ satisfies \eqref{eq:con5} since ${\delta}^*_{m+1}-{\delta}^*_{m}=\omega \left( g_{m+1}(p_{m+1})-g_{m+1}(p_{m}) \right)$.
Hence, ${\delta}_{m+1}^\ast$ in \eqref{eq:delta_k} is feasible, that is, it satisfies \eqref{eq:con3}--\eqref{eq:con5}.
Similarly, ${\delta}^*_{k}$ (for all $k \in\{ m+1,  \ldots, K\}$) in  \eqref{eq:delta_k} is feasible.

Now we prove that the monthly subscription fees in \eqref{eq:delta_1} and \eqref{eq:delta_k} form a best monthly fee assignment that can maximize the operator's profit.
The operator's profit collected from Bill APs is $\sum_{k\in\mathcal{B}}N_k\delta_k$.
Assume there is a monthly subscription fees assignment $\{\widetilde{\delta}_k\}$ such that $\sum_{k\in\mathcal{B}}N_k\widetilde{\delta}_k>\sum_{k\in\mathcal{B}}N_k{\delta}^*_k$.
So there is at least one fee $\widetilde{\delta}_i > {\delta}^*_i$.
To make the contract feasible, $\{\widetilde{\delta}_k\}$ must satisfy the following constraint according to Theorem 1:
\[ \widetilde{\delta}_i \leq \widetilde{\delta}_{i-1}+\omega (g_i(p_i)-g_i(p_{i-1})) .\]
Combining the above equation with \eqref{eq:delta_k}, we have:
\begin{align*}
& \widetilde{\delta}_{i-1} \geq \widetilde{\delta}_i - \omega (g_i(p_i)-g_i(p_{i-1})) \\
& ~~~~~~ > {\delta}^*_i - \omega (g_i(p_i)-g_i(p_{i-1}))={\delta}^*_{i-1} .
\end{align*}
Continuing the above process, we can finally obtain that $\widetilde{\delta}_{m}>{\delta}^*_{m}=\omega g_{m}(p_{m})-\beta_m$, which violates the IR constraint for type ${m}$.
Therefore, there does not exist any feasible $\{\widetilde{\delta}_k\}$ such that $\sum_{k\in\mathcal{B}}N_k\widetilde{\delta}_k>\sum_{k\in\mathcal{B}}N_k{\delta}^*_k$, which implies that the revenue is maximized under $\{{\delta}^*_k\}$.

Then we show that $\{{\delta}^*_k\}$ is the unique best monthly fees assignment.
Assume there is a $\{\widetilde{\delta}_k\} \neq \{{\delta}^*_k\}$ such that $\sum_{k\in\mathcal{B}}N_k\widetilde{\delta}_k =\sum_{k\in\mathcal{B}}N_k{\delta}^*_k$.
Without loss of generality, we assume that there is one fee $\widetilde{\delta}_j < \delta^*_j$.
It is easy to see that there must exist another fee $\widetilde{\delta}_i > {\delta}^*_i$.
Using the same argument, we can obtain that $\widetilde{\delta}_{m}>{\delta}^*_{m}=\omega g_{m}(p_{m})-\beta_m$.
Therefore, there does not exist any feasible $\{\widetilde{\delta}_k\} \neq \{{\delta}^*_k\}$ such that $\sum_{k\in\mathcal{B}}N_k\widetilde{\delta}_k=\sum_{k\in\mathcal{B}}N_k{\delta}^*_k$, which implies that the best monthly fees assignment $\{{\delta}^*_k\}$ is unique.
\end{proof}

\subsection{Proof of Lemma 5}

\begin{proof}
We prove this lemma by using the IC constraint.

We first prove that if $p_{k,l} > p_{i,j}$, then $\delta_{k,l} > \delta_{i,j}$.
For any APO characterized by $\eta_j ~(j\in\mathcal{M})$ and $\theta_i ~(i\in\BB_j)$, the following IC constraint must be satisfied:
\[ u_{i,j}(p_{i,j},\delta_{i,j};\Phi) \geq u_{i,j}(p_{k,l},\delta_{k,l};\Phi),\forall l \in\mathcal{M}, k\in \mathcal{B}_l, \]
which is
$$\omega_j(\Phi) g_i(p_{i,j})-\delta_{i,j}-\beta_{i,j}(\Phi) \geq \omega_j(\Phi)  g_i(p_{k,l})-\delta_{k,l} -\beta_{i,j}(\Phi).$$
This implies
\[ \delta_{k,l} - \delta_{i,j} \geq \omega_j(\Phi) ( g_i(p_{k,l})- g_i(p_{i,j}) ) > 0, \]
which follows the fact that $g_i(p)$ is an strictly increasing function with respect to $p$.

We now prove that if $\delta_{k,l} > \delta_{i,j}$, then $p_{k,l} > p_{i,j}$.
For any APO characterized by $\eta_l ~(l\in\mathcal{M})$ and $\theta_k ~(k\in\BB_l)$, the following IC constraint must be satisfied:
\[ u_{k,l}(p_{k,l},\delta_{k,l};\Phi) \geq u_{k,l}(p_{i,j},\delta_{i,j};\Phi),\forall j \in \mathcal{B}, \]
which is
$$\omega_l(\Phi) g_k(p_{k,l})-\delta_{k,l}-\beta_{k,l}(\Phi) \geq \omega_l(\Phi)  g_k(p_{i,j})-\delta_{i,j} -\beta_{k,l}(\Phi).$$
This implies
\[ g_k(p_{k,l})-g_k(p_{i,j}) \geq  \frac{\delta_{k,l} - \delta_{i,j}}{\omega_l(\Phi)} > 0. \]
Since $g_i(p)$ monotonically increases with $p$, we have
\[ p_{k,l} > p_{i,j} .\]
\end{proof}

\subsection{Proof of Lemma 6}

\begin{proof}

From the IC constraints for the APOs characterized by $\eta_l ~(l\in\mathcal{M})$ and $\theta_k ~(k\in\BB_l)$ and the APOs characterized by $\eta_j ~(j\in\mathcal{M})$ and $\theta_i ~(i\in\BB_j)$, we have
\begin{align}
& u_{k,l}(p_{k,l},\delta_{k,l};\Phi) \geq u_{k,l}(p_{i,j},\delta_{i,j};\Phi), \notag\\
& u_{i,j}(p_{i,j},\delta_{i,j};\Phi) \geq u_{i,j}(p_{k,l},\delta_{k,l};\Phi), \notag
\end{align}
which are equivalent to
\begin{align}
& \omega_l(\Phi) g_k(p_{k,l})-\delta_{k,l}-\beta_{k,l}(\Phi) \geq \omega_l(\Phi)  g_k(p_{i,j})-\delta_{i,j} -\beta_{k,l}(\Phi), \notag\\
& \omega_j(\Phi) g_i(p_{i,j})-\delta_{i,j}-\beta_{i,j}(\Phi) \geq \omega_j(\Phi)  g_i(p_{k,l})-\delta_{k,l} -\beta_{i,j}(\Phi). \notag
\end{align}
Combining the above two inequations, we have
\begin{equation}\label{eq:proof}
\omega_l(\Phi) (g_k(p_{k,l})-g_k(p_{i,j})) \geq \omega_j(\Phi) ( g_i(p_{k,l})- g_i(p_{i,j}) ).
\end{equation}

If $j=l$ and $\theta_k>\theta_i$, \eqref{eq:proof} is equivalent to
\[g_k(p_{k,l})-g_k(p_{i,l}) \geq  g_i(p_{k,l})- g_i(p_{i,l}),\]
which implies that $p_{k,l}>p_{i,l}$.

If $i=k$ and $\eta_l > \eta_j$, \eqref{eq:proof} is equivalent to
\[\omega_l(\Phi) (g_k(p_{k,l})-g_k(p_{k,j})) \geq \omega_j(\Phi) ( g_k(p_{k,l})- g_k(p_{k,j}) ),\]
which implies that $p_{k,l} \geq  p_{k,j}$.

If $\eta_l < \eta_j$ and $\theta_k>\theta_i$, \eqref{eq:proof} is equivalent to
\[  g_k(p_{k,l})-g_k(p_{i,j}) >  g_i(p_{k,l})- g_i(p_{i,j})  ,\]
which implies that $p_{k,l}>p_{i,j}$.

\end{proof}

\subsection{Proof of Lemma 7}

\begin{proof}
We prove the lemma by the contradiction principle.
%
Suppose the APOs characterized by $\eta_l ~(l\in\mathcal{M})$ and $\theta_k$ choose to be Linus, but the APOs characterized by $\eta_l ~(l\in\mathcal{M})$ and $\theta_i ~(i<k)$ choose to be Bills.

From the IC constraint for the (Linus) APO characterized by $\eta_l$ and $\theta_k$, we have
\[ u_{k,l}(p_{i,l},\delta_{i,l};\Phi) \leq u_{k,l}(0,0),\]
equivalently,
\[ \mu(\Phi) g_k(p_{i,l})-\delta_{i,l}-\frac{1-\eta_l}{N} \nu(\Phi)\leq 0 .\]

For the (Bill) APO characterized by $\eta_l$ and $\theta_i ~(i<k)$, we have
\begin{align*}
u_{i,l}(p_{i,l},\delta_{i,l};\Phi) &= \mu(\Phi) g_i(p_{i,l})-\delta_{i,l}-\frac{1-\eta_l}{N}\nu(\Phi) \\
& < \mu(\Phi) g_k(p_{i,l})-\delta_{i,l}-\frac{1-\eta_l}{N}\nu(\Phi) \\
& = u_{k,l}(p_{i,l},\delta_{i,l};\Phi) \leq 0,
\end{align*}
which contradicts with the IR constraint for the (Bill) APO $\eta_l$ and $\theta_i ~(i<k)$.

Therefore, for a feasible contract $\Phi$, there exists a critical APO type $m_l\in\{1,2,\ldots,K+1\}$ for each $\eta_l ~(\forall l\in\mathcal{M})$ such that $k \in \mathcal{L}_l$ for all $k < m_l$, and $k \in \mathcal{B}_l$ for all $k \geq m_l$.
\end{proof}

\subsection{Proof of Lemma 8}

\begin{proof}
We prove this lemma by contradiction.
Assume to the contrary that $\exists l>j$ such that $m_l > m_j$.
Then there exists $k\in\K$ such that $k \in \mathcal{L}_l$ (the type-$(k,l)$ APOs choose to be Linus), but $k \in \mathcal{B}_j$ (the type-$(k,j)$ APOs choose to be Bills).

From the IC constraint for the type-$\{k,j\}$ (Bill) APO (characterized by $\eta_j$ and $\theta_k$), we have
\[
u_{k,j}(p_{k,j},\delta_{k,j};\Phi) = \mu g_k(p_{k,j})-\delta_{k,j}- \frac{1-\eta_j}{N} \nu \geq 0.
\]

Since $l>j$, we have $\eta_l > \eta_j$.
Then for the type-$\{k,l\}$ Linus APO (characterized by $\eta_l$ and $\theta_k$), if he chooses to be Bill, we have
\begin{align*}
u_{k,l}(p_{k,j},\delta_{k,j};\Phi) &= \mu g_k(p_{k,j})-\delta_{k,j}-\frac{1-\eta_l}{N} \nu \\
&> \mu g_k(p_{k,j})-\delta_{k,j}-\frac{1-\eta_j}{N} \nu \geq 0,
\end{align*}
which contradicts with the IC constraint for the type-$\{k,l\}$ Linus APO.

Hence, we have $m_l \leq m_j, \forall l,j\in\mathcal{M}, l>j$.
\end{proof}

\end{document}